\documentclass[9pt,twocolumn]{article}
\usepackage{arxiv}
\usepackage[utf8]{inputenc}
\usepackage{amsmath}
\usepackage{authblk} % for affiliations
\usepackage{amssymb}
\usepackage{graphicx}
\usepackage{multirow}
\usepackage{multicol} % allows switch two 2-column on same page (\begin{multicols}{2}....)
\usepackage{hyperref}
\usepackage{balance}
\usepackage{color}
\usepackage[resetlabels]{multibib}
\newcites{SI}{References}

\usepackage{xcolor} % for \textcolor{red}{sadflkasjdflasjk}

\newcommand*\colvec[3][]{
    \begin{pmatrix}\ifx\relax#1\relax\else#1\\\fi#2\\#3\end{pmatrix}
}   % \colvec{a}{b} = (a, b), \colvec[a]{b}{c} = (a, b, c)
\usepackage{chngcntr}
\counterwithout{equation}{section}

\title{Collective predator evasion: Putting the criticality hypothesis to the test}
\author[a,b]{Pascal P. Klamser}
\author[a,b]{Pawel Romanczuk}
\affil[a]{Department of Biology, Institute for Theoretical Biology, Humboldt‐Universität zu Berlin, 10115 Berlin, Germany}
\affil[b]{Bernstein Center for Computational Neuroscience, 10115 Berlin, Germany}

\begin{document}
\onecolumn

\maketitle
%\tableofcontents

\begin{abstract}
According to the \emph{criticality hypothesis}, collective biological systems should operate in a special parameter region, close to so-called critical points, where the collective behavior undergoes a qualitative change between different dynamical regimes. Critical systems exhibit unique properties, which may benefit collective information processing such as maximal responsiveness to external stimuli. Besides neuronal and gene-regulatory networks, recent empirical data suggests that also animal collectives may be examples of self-organized critical systems. However, open questions about self-organization mechanisms in animal groups remain: Evolutionary adaptation towards a group-level optimum (group-level selection), implicitly assumed in the ``criticality hypothesis'', appears in general not reasonable for fission-fusion groups composed of non-related individuals. Furthermore, previous theoretical work relies on non-spatial models, which ignore potentially important self-organization and spatial sorting effects.
Using a generic, spatially-explicit model of schooling prey being attacked by a predator, we show first that schools operating at criticality perform best. However, this is not due to optimal response of the prey to the predator, as suggested by the ``criticality hypothesis'', but rather due to the spatial structure of the prey school at criticality. Secondly, by investigating individual-level evolution, we show that strong spatial self-sorting effects at the critical point lead to strong selection gradients, and make it an evolutionary unstable state. Our results demonstrate the decisive role of spatio-temporal phenomena in collective behavior, and that individual-level selection is in general not a viable mechanism for self-tuning of unrelated animal groups towards criticality.

\end{abstract}
\twocolumn
% \begin{multicols}{2}
\section{Introduction}

Distributed processing of information is at the core for the function of many complex systems in biology, such as neuronal networks \cite{Rubinov2010}, genetic regulatory networks \cite{Vijesh2013} or animal collectives \cite{Miller2013, Strandburg-Peshkin2013}. Based on ideas initially developed in statistical physics and theoretical modeling it has been conjectured that such living systems operate in a special parameter region, in the vicinity of so-called critical points (phase transitions), where the system's macroscopic dynamics undergo a qualitative change, and various aspects of collective computation become optimal \cite{Packard1988, Bak1989, Langton1990, Bak1993, Kinouchi2006, Mora2011, Hidalgo2014}. 
In recent years some empirical support for the ``criticality hypothesis'' has been obtained from analysis of neuronal dynamics \cite{Beggs2012, Mora2011, Munoz2018}, gene regulatory networks \cite{krotov2014morphogenesis,daniels2018criticality}, and collective behaviors of animals \cite{Gelblum2015,Feinermann,Attanasi2014,Bialek2013,Daniels2017,Ginelli2015a}. 
This evidence is often based on observation of characteristic features of critical behavior, such as power-law distribution or diverging correlation lengths in spatial systems. However these observations could in principle have different origins  \cite{Beggs2012, Lorimer2015, Reed2002, touboul2017power}. 
Therefore, more convincing support for the ``criticality hypothesis'' can be obtained through additional identification of proximate mechanisms enabling biological systems to self-organize towards criticality. 
In neuronal systems, synaptic plasticity has been shown to provide such a  mechanism  \cite{bornholdt2000topological,meisel2009adaptive,ma2019cortical}. 
For genetic regulatory networks, similar mechanisms based on network rewiring have been proposed  \cite{liu2006emergent,macarthur2010microdynamics}. Using an information-theoretic framework Hidalgo et al. \cite{Hidalgo2014} have shown that (coupled) binary networks evolve towards the critical state in heterogeneous environments. However, in their model already a single unit (individual) can exhibit a phase-transition and thus tunes itself individually to criticality. In addition, they assumed idealized random interaction networks between the agents. Thus, open questions remain whether evolutionary, individual-level adaptations is a possible self-tuning mechanism for (i) biological collectives, where phase transitions are purely macroscopic phenomena, and (ii) animal groups characterized by spatial, dynamic interaction networks. In general, if collective computation becomes optimal at a phase transition, a purely macroscopic phenomenon defined only at the group-level, then adaptation based on global fitness should be able to tune the system towards criticality. Therefore, at first glance Darwinian evolution appears a viable mechanism for emergence of self-organized criticality only for complex systems within a single individual, e.g. in the context of neuronal or genetic networks, or in collectives of closely related individuals such as eusocial insects \cite{Gelblum2015}. In multi-agent systems group-level and individual-level evolutionary optima are often different\cite{Torney2015, Brush2016}, leading to so-called social dilemmas emerging in a broad range of multi-agent evolutionary game theoretic problems. In the context of animal groups consisting of non-related individuals, this questions individual-level adaptation as a proximate mechanism for self-tuning of collective systems to criticality as a potential group level optimum. Here multi-level selection has been proposed to address some related fundamental problems in the evolution of collective behavior \cite{Wilson1975, Wilson1997}. However, it has been recently shown that even under strong group-level selection, as long as individual-level selection plays a non-negligible role, multi-level selection will also result in evolution of sub-optimal collective behaviors \cite{cooney2019replicator,cooney2020analysis}. 

Whereas few empirical studies report signatures of criticality in collective animal behavior \cite{Bialek2013, Attanasi2014, Daniels2017}, most support for the criticality hypothesis in this context comes from mathematical models.
For example, in agent-based simulation of fish schools it has been shown that at a critical point the collective state is influenced strongest by single or few individuals \cite{Calovi2015}, or that collective response to external time-varying signals becomes maximal in idealized lattice models of flocks \cite{Vanni2011}.
However, dynamical animal groups differ from lattice models \cite{Vanni2011, Chicoli2016} due to their dynamical neighborhood which may induce self-sorting of individuals according to their individual behavioral parameters \cite{Couzin2002, Hemelrijk2005, JamieWood2010}. This, in turn, has likely direct evolutionary consequences as for example predators may attack certain swarm regions more frequently \cite{KRAUSE1994, Bumann1997, Handegard2012}.

Throughout this work, criticality or critical point will refer to the directional order-disorder transition, a prominent phase transition in statistical physics and collective behavior \cite{Vicsek1995}:
An initially disordered swarm, where the social coordination is weak compared to noise, shows spontaneous onset of orientational order, if the directional alignment (coupling strength) is increased beyond a critical parameter (critical point): The group starts to move collectively along a common "consensus" direction. A further increase of alignment results in highly ordered (polarized) schools \cite{Grossmann2012}.
This transition is characterized by a so-called spontaneous symmetry breaking: In disordered swarms there is no distinguished direction in space. In the ordered state, this symmetry is broken through the emergence of an average heading direction of the school, which allows to distinguish front, back and sides of the group.

We explore the criticality hypothesis in the context of spatially-explicit predator-prey dynamics, where coordinated collective behavior of the prey is believed to entail evolutionary benefits to individuals within the group \cite{Krause2002}.
In particular, we use an agent-based model of grouping and coordinating prey \cite{Harpaz2017, Katz2011a, Calovi2018, Herbert-Read2017, Couzin2002, Hemelrijk2005}, and analyze the role of the spatial structure of the group, its dynamical response and the individual-level selection by applying an evolutionary algorithm \cite{Wood2007, Olson2012, Olson2016, Hein2015, Guttal2010, Monk2018, VanDerPost2015}.

We show that the group-level behavior becomes optimal at criticality with respect to two measures: We observe i) maximal directional-information transfer between neighbors, and ii) minimal predator capture rates at criticality. However, a detailed analysis reveals that the capture rate, as a relevant measure of evolutionary fitness, becomes minimal only due to the dynamical structure of the collective at criticality, independent of the direct response of individuals to the predator, and thus independent of information propagation within the school.
Furthermore, through evolutionary simulations with individual-level selection, we show that the critical point is an evolutionary highly unstable state. This evolutionary instability can be linked to strong selection due to phenotypic-sorting with respect to the broken symmetry of the collective state. Finally, the observed evolutionary stable strategies (ESS) result from individual prey agents balancing the influence of social and private information on their movement response.

\begin{figure}[h]
  \includegraphics[width=\linewidth]{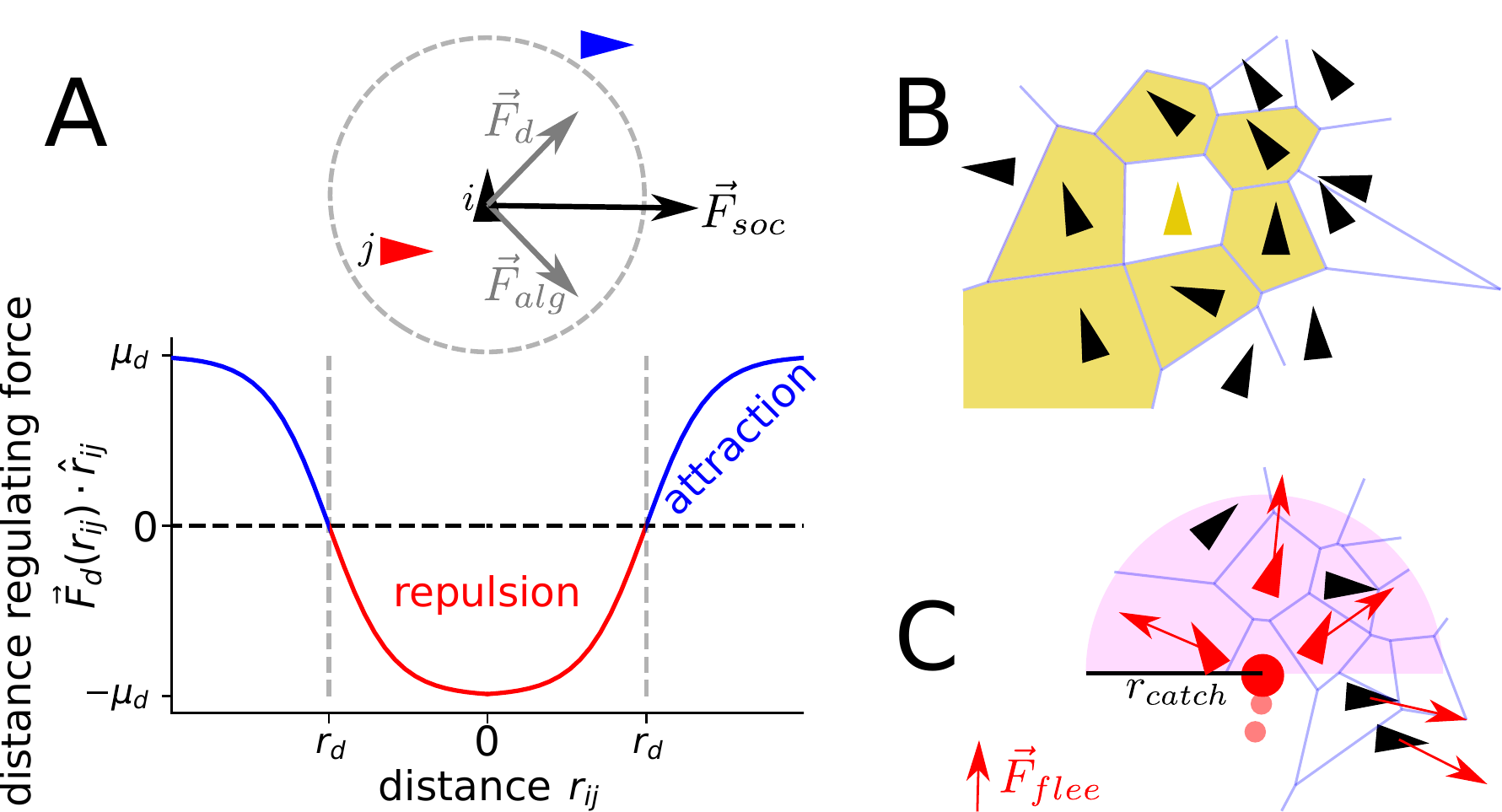}
    \caption{
        \textbf{Social forces, interaction network and predator-prey interaction.}
        Schematic illustration of social forces (\textbf{A}), the Voronoi interaction network (\textbf{B}) and the predator prey interactions (\textbf{C}).
        (\textbf{A}) The social force $\vec{F}_{soc}$ acts on the focal agent (black triangle) and is a combination of alignment $\vec{F}_{alg}$ and distance regulation $\vec{F}_d$ to its interaction partners.
        The alignment is proportional to the sum of the velocity differences $F_{i, alg}\propto \sum_j \vec{v}_{ji}$ with $\vec{v}_{ji} = \vec{v}_j - \vec{v}_i$ and thus not parallel to the neighbors mean velocity but tends to minimize the velocity difference.
        The distance regulating force $F_d$ is a continuous version of a two zone model, i.e. the focal agent is repelled from neighbors that are closer (\textcolor{red}{red triangle/line}) than the preferred distance $r_d$ (\textcolor{gray}{gray dashed circle/line}) and attracted to those farther away (\textcolor{blue}{blue triangle/line}). 
        (\textbf{B}) A focal prey agent (yellow triangle) interact with it's nearest Voronoi neighbors (black triangles in yellow cells).
        (\textbf{C}) The predator (\textcolor{red}{red point}) pursues the weighted mean direction of the targets (\textcolor{red}{small red triangles}), which are the frontal Voronoi neighbors. Their weight is proportional to their probability of capture, which decreases linear with distance and is zero for $r\geq r_{catch}$ (\textcolor{magenta}{magenta semicircle}).
        All Voronoi neighbors of the predator flee with a repulsive force $\vec{F}_{flee}$ (\textcolor{red}{red arrows}).
        }
    \label{fig:modelExplain}
\end{figure}

\section{Results}

\subsection{Agent based model of predator-prey interactions}

We consider a simple, yet generic agent-based model of schooling prey attacked by a predator. For simplicity we assume initially that the prey agents move with fixed speed $v_0$ and change their direction according to social forces (Fig~\ref{fig:modelExplain}A): they tend to keep a preferred distance to, and align (alignment strength $\mu_{alg}$) their velocities with the first shell of nearest neighbors, defined by a Voronoi tessellation (Fig~\ref{fig:modelExplain}B) \cite{Strandburg-Peshkin2013, Ballerini2008a}.
A distance regulating represents a continuous version of a two zone model, i.e. repulsion at short distances and an attraction zone at large distances with a "preferred" (equilibrium) inter-individual distance $r_d$ (Fig~\ref{fig:modelExplain}A).
Randomness in the movement of individuals due to unresolved internal decisions or environmental noise is modeled as fluctuations in the heading of the agents (angular noise with intensity $D$).
The prey responds with a flee-force with strength $\mu_{flee}$ to a predator within its Voronoi neighbors (Fig~\ref{fig:modelExplain}C).

The predator moves with a fixed speed $v_p$ which is larger than the preys (here $v_p=2v_0$) and its direction changes towards the weighted mean direction of its frontal nearest prey, which represent possible targets (Fig~\ref{fig:modelExplain}C).
The weight corresponds to the catch-probability of each target, which decreases linearly with distance until it equals zero at a distance larger than the catch-radius.
If the predator launches an attack, with attack rate $\gamma_a$, it selects equally likely among the possible targets and captures it according to the targets catch-probability.
The predator is initiated outside the prey collective with a distance slightly above the capture-radius and a velocity vector oriented towards the center of mass of the prey school.

In evolutionary simulations for each generation we perform $N_r$ independent runs with different initial conditions for $N$ agents, each with its behavioral phenotype defined by the evolvable social force parameter (alignment strength $\mu_{alg}$). Fitness of a prey agent is defined through the negative number of deaths of this agent aggregated over the $N_r$ independent runs.
The behavioral phenotypes, i.e. social force parameters, of the next generation are selected via fitness-proportionate selection (roulette-wheel-algorithm)\cite{Guttal2010,guttal2012cannibalism,lipowski2012roulette} with mutations implemented through addition of Gaussian-distributed noise on the selected behavioral parameter. See methods for model details.

\subsection{Collective information transfer and responsiveness}

\begin{figure}[h]
  \includegraphics[width=\linewidth]{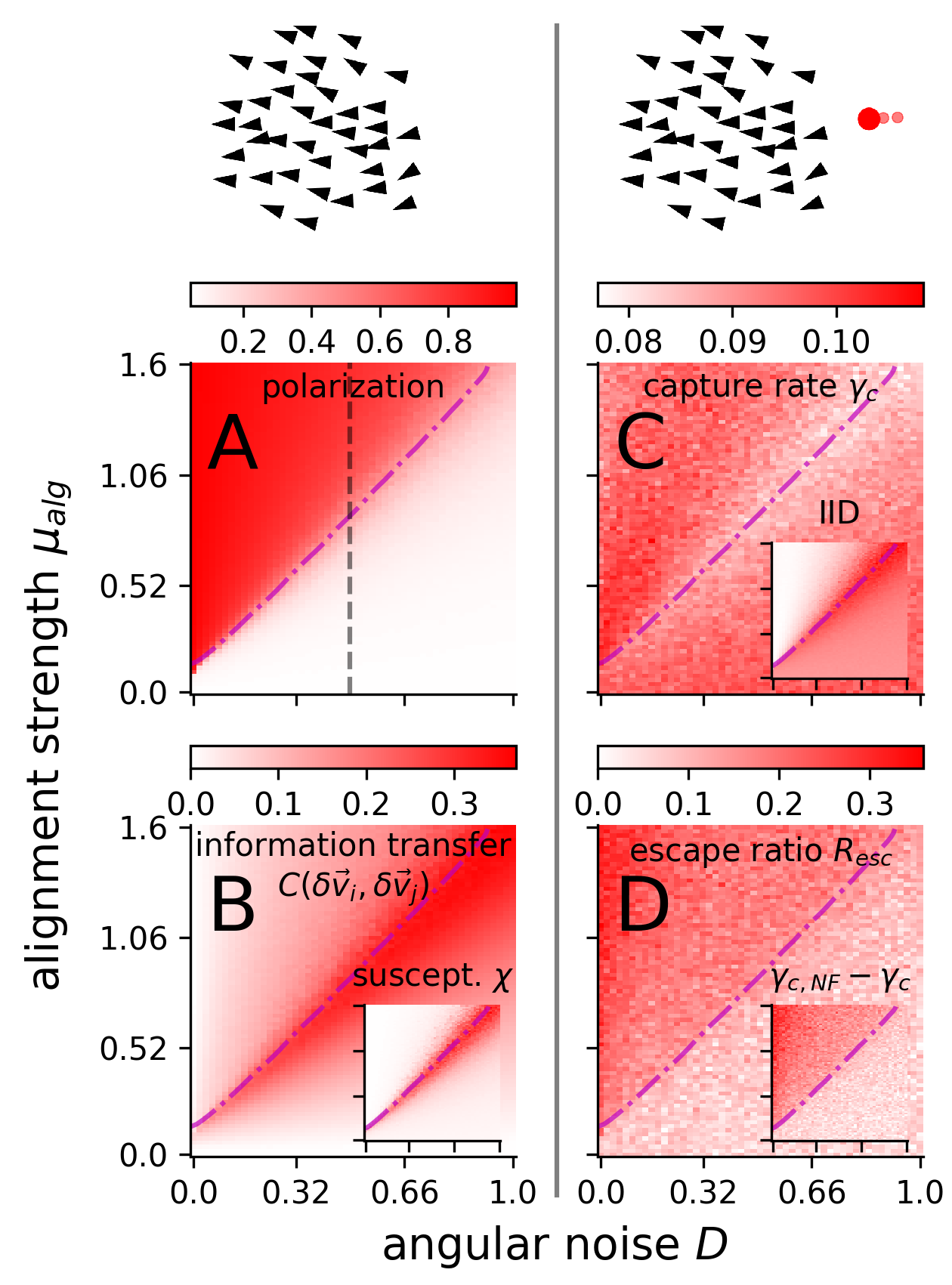}
    \caption{
        \textbf{Group optimum.} Predation independent (\textbf{A, B}) and dependent (\textbf{C, D}) group measures.
        (\textbf{A}) Polarization $\Phi$. The dashed vertical line marks the angular noise of $D=0.5$ used in the evolutionary runs.
        (\textbf{B}) Directional information transfer $C(\delta \vec{v}_i, \delta \vec{v}_j)$, estimated via the correlation of velocity fluctuations between interacting agents, peaks at the transition. Inset: Susceptibility, estimated via polarization fluctuations.
        (\textbf{C}) Collective anti-predator performance quantified by the capture rate, which is strongly anti-correlated with the inter-individual distance $R=-0.69$ (IID, inset \textbf{C}).
        (\textbf{D}) Escape ratio $R_{esc}=1 - \gamma_c / \gamma_{c, NF}$. Inset: Difference between capture rates in schools of non-fleeing $\gamma_{c, \text{NF}}$ and fleeing $\gamma_c$ agents.
        In all panels: the disorder-order transition is indicated by a the dash-dotted magenta line.
        Each parameter point corresponds to an average over $N_s=40$ simulations, each with $N=400$ agents attacked for $T_{simu}=120$ time units after an equilibration time of $T_{eq}=200$.
        For all insets(B, C, D): colorbars are shown separately in Fig.~\ref{fig:SI_GrpOptimaInsets}.
        }
    \label{fig:grpOptima}
\end{figure}

We first investigate whether operating at the order-disorder transition leads to optimal response of the prey school to the predator. Here, polarization $\Phi$, i.e. the normalized average velocity of the group, is the relevant order parameter quantifying the amount of orientational order in the system: For large, disordered systems $\Phi$ is close to zero, while in completely ordered systems with all agents moving in the same direction it approaches $1$ (see methods). It increases with the strength of alignment $\mu_{alg}$ and decreases with the intensity of angular noise $D$ (\ref{mov:1}~Video) in a non-linear fashion: It remains small ($\Phi \approx 0$) throughout most of the disordered regime, before showing the steepest increase in orientational order in the vicinity of the critical point, and finally asymptotically approaching $\Phi=1$.
Both behavioral parameters, $\mu_{alg}$ and $D$ can be used as control parameters for crossing of the critical line (diagonal magenta line Fig \ref{fig:grpOptima}A) between the disordered state (low $\mu_{alg}$, high $D$) and the ordered state (high $\mu_{alg}$, low $D$).

A simple and intuitive measure of responsiveness of such a collective system to (local) perturbations is the average pair-wise correlation of velocity fluctuations $C_{ij}=C(\delta \vec{v}_i,\delta \vec{v}_j)$ between interacting agents (see methods). Here, $\delta  \vec{v}_i= \vec{v_i} - \langle \vec{v} \rangle$ is the deviation of the velocity of agent $i$ from the average school velocity $\langle \vec{v} \rangle$. $C_{ij}$ can be interpreted as a simple measure of directional information transfer between neighboring agents $i$ and $j$:  If agent $i$ deviates from the average group direction due to a perturbation, large values of $C_{ij}$ indicates that agent $j$ to a large degree is "copying" this velocity deviation or vice versa.
% On the other hand, $C_{ij}\approx 0$ indicates that both agents perform random turns independent of each other, with vanishing directional information exchange.

The velocity fluctuation correlation $C_{ij}$ is closely related to the susceptibility $\chi$, which in statistical physics quantifies the degree of responsiveness of the system to perturbations, and may become maximal at criticality. 
It can be defined analogous to magnetic susceptibility in physics \cite{Calovi2015, Cavagna2017} (see methods).

Both measures, $C_{ij}$ and $\chi$, show a peak at the transition between order and disorder (see Fig \ref{fig:grpOptima}A, B) in line with predictions of the ``criticality hypothesis'' \cite{Munoz2018}. In terms of directional information transfer, i.e. the directional responsiveness to perturbation, it appears to be optimal for the collective to operate at criticality.

\subsection{Fitness relevant performance measure}

The validity of the above variables from a statistical physics point of view relies on the assumptions of homogeneity and temporal stationarity of the external field, which is not fulfilled in our predator-prey scenario: predator perturbation represents a strongly local, nonlinear perturbation. As a biologically relevant measure, independent of these assumptions, we use directly the predator capture rate $\gamma_c$, computed as number of prey captured per time unit. In agreement with the previous response measures, we find that the capture rate also exhibits a distinct minimum at the critical point (Fig \ref{fig:grpOptima}C).

However, varying the behavioral parameters of the prey (alignment strength or noise) not only changes the polarization of the school and the information transfer capability but it also affects the spatial structure of the school (\ref{mov:1}~Video, \ref{mov:2}~Video), e.g. the average inter-individual distance (IID) or the shape of the school. Our results show that structural properties of the prey school correlate strongly with the capture-rate, e.g. the inter-individual distance (inset Fig \ref{fig:grpOptima}C) with $C(\gamma_c, IID) \approx -0.69$.
Thus, the reduced capture rate may be potentially related to changes in the structure of the school at criticality.
To distinguish whether structure or information transfer is responsible for the optimal performance of the group at the critical point, we simulated for each predator attack a non-fleeing prey school (flee strength $\mu_{flee}=0$) as a control. This non-responsive control school is identical to the responsive school in all the remaining parameters and in its positions and velocities at the time of predator appearance (see \ref{mov:3}~Video).
The capture-rate of the non-fleeing prey $\gamma_{c, NF}$ depends only on the self-organized structure of the school. 
We compare the responsive and control school via two measures: (i) the simple difference between both capture rates $\gamma_{c, NF} - \gamma_{c}$ and (ii) the escape ratio $R_{esc}$, which is more robust to fluctuations (see methods) and is defined as the fraction of surviving responsive prey, which would have been captured if they would not flee.
Interestingly both measures show no peak at the transition but a continuous increase with alignment strength (Fig \ref{fig:grpOptima}D) suggesting that the predator-response improves towards the ordered phase if we control for the differences in the self-organized spatial structure (compare column $\mu_{alg}=1$ with $\mu_{alg}=2$ in \ref{mov:2}~Video).

These results demonstrate that the direct cause of the optimal collective performance (minimal capture rate) is the dynamical structure, as a "passive" component, and surprisingly not the maximal responsiveness at criticality (see SI~Sec.~\ref{sec:cri_suscVsPredResp} for theoretical reasoning on differences between susceptibility and predator response).

\subsection{Evolution of coordinated escape}

\begin{figure*}[h]
    \includegraphics[width=1\textwidth]{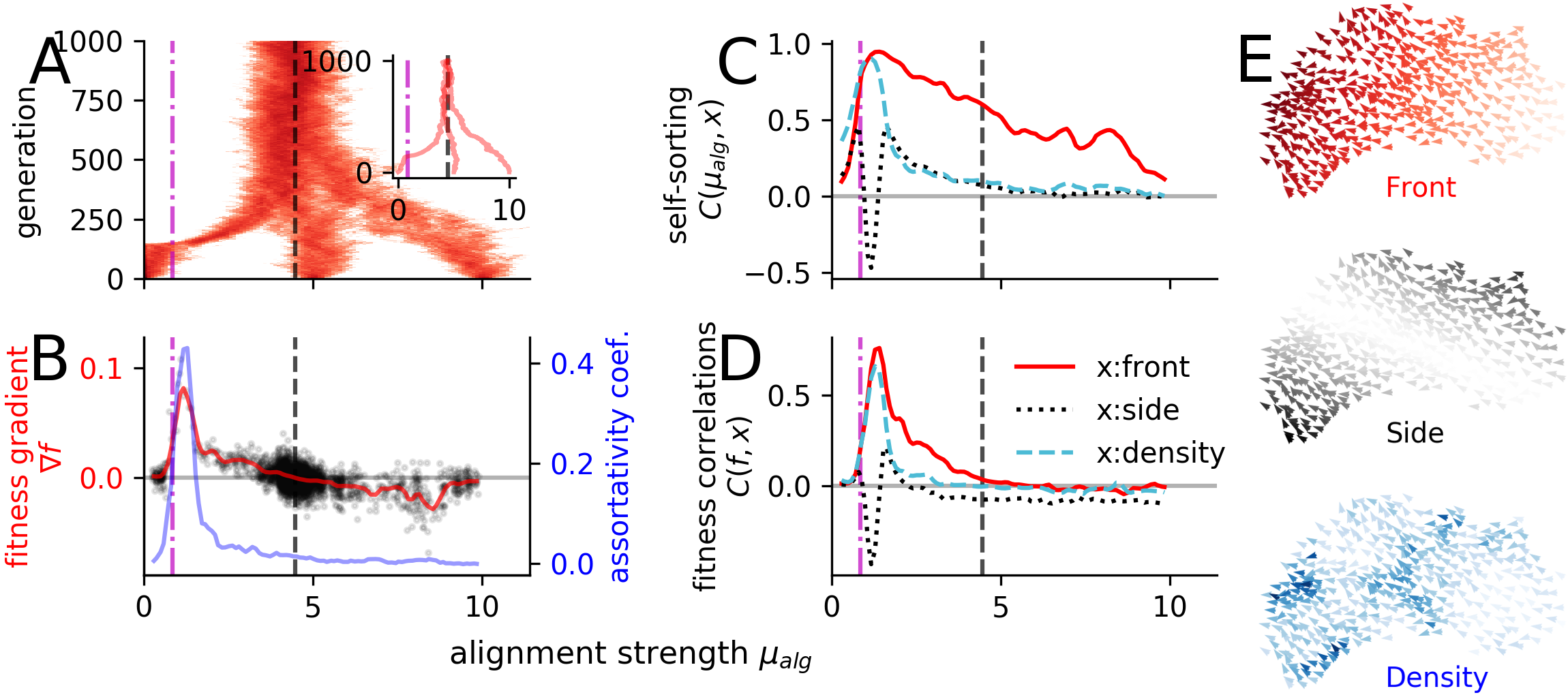}
    \caption{\textbf{Evolution under predation.}
        (\textbf{A}) Overlay of three independent evolutionary runs starting at $\langle \mu_{alg} \rangle = [0, 5, 10]$ over 1000 generations. The behavioral phenotype is determined only by the alignment strength as the evolving parameter. The predator attacks from random initial directions for $T_{simu}=120$.
        The inset shows the evolution of the population mean alignment parameter $\langle \mu_{alg} \rangle$ of the three different evolutionary runs.
        (\textbf{B}) Assortativity coefficient (\textcolor{blue}{blue line}) and smoothed fitness gradient $\nabla f$ (\textcolor{red}{red line}).
        The evolutionary stable state is defined by the zero crossing of the fitness gradient and represented as a vertical dashed black line.
        Black dots are the non-averaged fitness gradients for each generation (see methods).
        (\textbf{C}) Self-sorting measured as correlation $C(\mu_{alg}, x)$ between the individual alignment strength $\mu_{alg}$ and variables quantifying its (spatial) location within the school: front-back position (\textcolor{red}{red}) and side-center position (black) and local density (\textcolor{blue}{blue}).
        (\textbf{D}) Correlation $C(f, x)$ of individual fitness with the average relative spatial positions.
        (\textbf{E}) Simulation snapshot illustrating the location variables: front-back position (\textcolor{red}{red}) and side-center position (black) and local density (\textcolor{blue}{blue}).
        In all panels: the vertical dash-dotted magenta line marks the order-disorder transition and the vertical dashed black line the evolutionary stable state.
    }
    \label{fig:EvoOptima}
\end{figure*}

The group-optimum at criticality with respect to prey-survival, does not need to coincide with the evolutionary stable state (ESS) with respect to evolutionary adaptations at the individual level.
To explore whether the transition region is favored by individual-level adaptation, we let the individual alignment strength $\mu_{alg}$ evolve over 500 generations, while keeping the angular noise constant ($D=0.5$: vertical line Fig \ref{fig:grpOptima}A). We repeat the evolutionary simulations from different initial conditions: below ($\langle\mu_{alg}\rangle=0$), above ($\langle\mu_{alg}\rangle=5$) and far above ($\langle\mu_{alg}\rangle=10$) the transition ($\mu_{c, alg}\approx0.9$).
To ensure that the evolution ends at the ESS we compute the fitness gradient which represent the strength of the selection pressure at a specific mean alignment strength (see methods). Assuming a monomodal phenotype distribution, as observed in our evolutionary runs, a change in sign of the fitness gradient marks the location of the ESS.
All three initiations end in the ordered region far above the critical point (Fig \ref{fig:EvoOptima}A) and fluctuate around $ESS(\mu_{alg})\approx4.4$ (vertical dashed line Fig \ref{fig:EvoOptima}B).
Thus, the transition region is not an attractor of the evolutionary dynamics. On the contrary, it is a highly unstable point with fast evolutionary dynamics due to particularly strong selection pressure at criticality. The fitness gradient peaks shortly above the transition in the ordered phase (Fig \ref{fig:EvoOptima}B), with evolutionary dynamics pushing the system out of the transition region towards stronger alignment.

A possible driver of this maximal selection pressure is self-sorting, i.e. the tendency of individuals to sort according to their behavioral parameters along specific spatial dimensions of the school, e.g. front-back or side-center, or in regions of higher or lower density (Fig \ref{fig:EvoOptima}C) \cite{Couzin2002}. We can quantify self-sorting through the Pearson correlation coefficient between the alignment strength (social phenotype) of an agent and variables quantifying its location within the school (see methods). Another measure of self-sorting is the amount of assortative mixing in the school as quantified by the assortativity coefficient (see methods). Assortativity (Fig \ref{fig:EvoOptima}B) as well as other self-sorting measures (Fig \ref{fig:EvoOptima}C) exhibit extrema which coincide with the fitness gradient peak.
Note that a strong assortative mixing is equivalent to the formation of spatially coherent sub-groups within the school with similar behavioral parameter.
In this context a peak in fitness gradient close to transition suggests that sub-groups with stronger alignment, thus better directional coordination, actively or passively perform better at avoiding capture. An increase in the escape ratio $R_{esc}$ with increasing alignment close to criticality (see Fig \ref{fig:grpOptima}D) suggest an enhanced active avoidance.
 However, also passive effects appear to play an important role since the correlation between the fitness of a prey and its relative position becomes maximal in the same parameter region (Fig \ref{fig:EvoOptima}D).
 One specific mechanism of passive avoidance is the dilution effect \cite{Krause2002} caused by local density differences correlating with behavioral phenotypes. Stronger aligning individuals form denser regions within the prey school (density-sorting Fig \ref{fig:EvoOptima}B). As a consequence they have a systematically smaller domain of danger \cite{james2004geometry} and are thus less frequently attacked by the predator.

It is possible to disentangle passive, structural effects from an active response, by setting the flee-strength to zero. This results in a significantly smaller, yet finite, fitness-gradient-peak at the transition (Fig.~\ref{fig:SI_EvoVaryFlee_detail}, panel H). This suggests that both, the structural, passive selection and the different active avoidance behavior of different phenotypes contribute to the strong selection pressure at criticality.

We note that the sudden increase in self-sorting at the transition is due to a coupled symmetry breaking. 
At the order-disorder transition the directional symmetry is broken and the school "agrees" on a common movement direction. 
This also breaks the symmetry between relative locations within the school.
For example in the disordered phase every edge position is equivalent, but with the emergence of the common movement direction the sides and rear of the school become structurally different from the front.
This can be clearly seen in the comparison of the correlations of individual alignment strength and specific relative spatial positions within the school ("side-sorting" versus "front-sorting"): Below the transition the corresponding curves become indistinguishable, whereas above at the transition they start to deviate and show different behavior with increasing alignment strength (Fig \ref{fig:EvoOptima}C).

\subsection{ESS: Balancing benefits and costs of social information}

Despite the importance of self-sorting for the maximal selection pressure at the transition, it does not provide an explanation for the observed location of the ESS. More specifically, it can not explain the negative fitness gradient for strong alignment $\mu_{alg}>ESS(\mu_{alg})\approx 4.4$.
In this regime either the self-sorting is negligible, as for side- and density-sorting (Fig \ref{fig:EvoOptima}B), or the relative location has no effect on the individual fitness, as observed along the front-back dimension (Fig \ref{fig:EvoOptima}E).
If the ESS is not determined by the structural self-organization of the school, it has to originate from individuals avoiding the predator better than others.
Please note that avoidance does not only mean to escape if targeted but also to avoid becoming a target.
In this case the ESS has to depend on the flee-strength $\mu_{flee}$ as the main parameter tuning the strength of individual predator response. 

\begin{figure}
    \includegraphics[width=\linewidth]{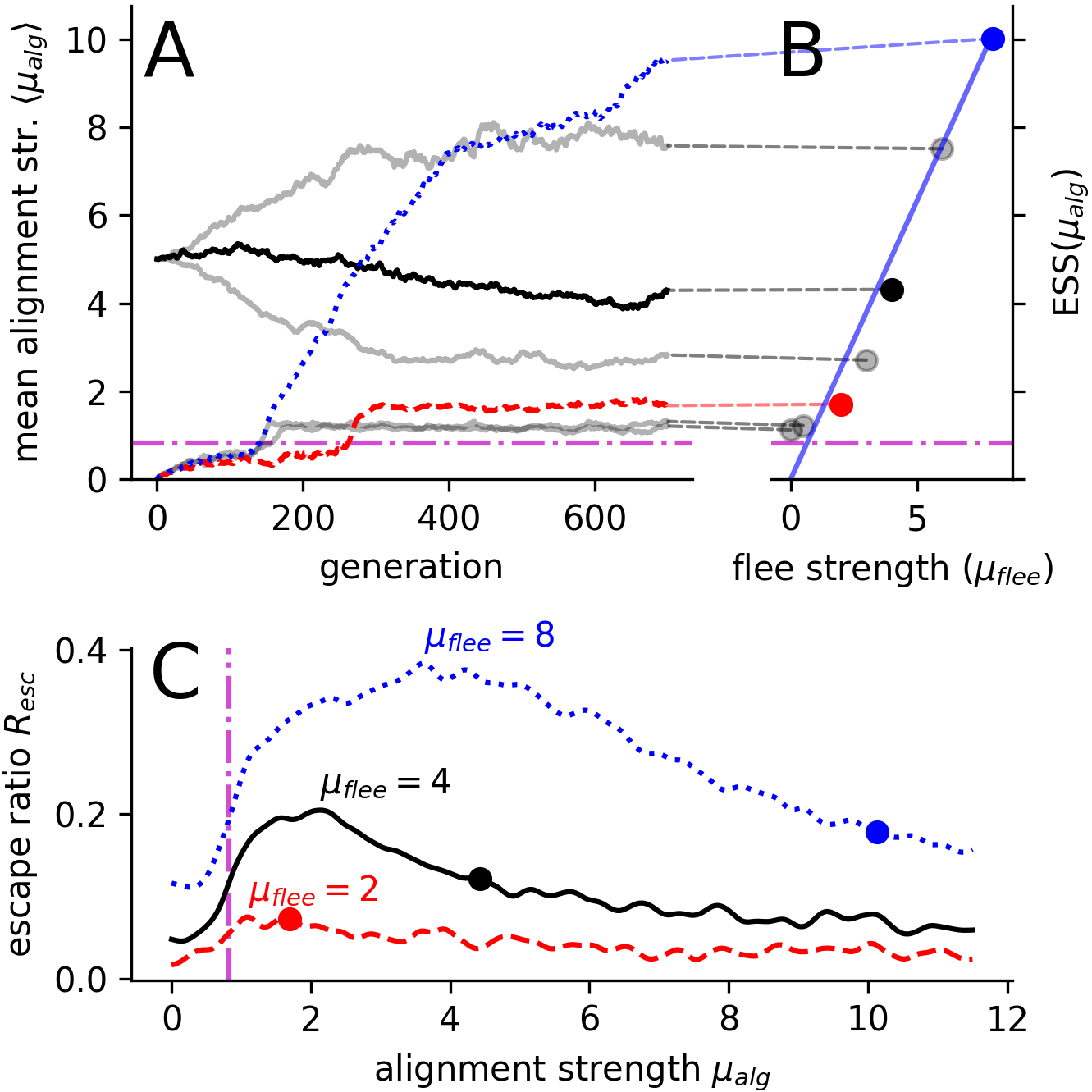}
    \caption{\textbf{Evolution for different flee strengths $\mu_{flee}$.} 
        (\textbf{A}) Sample evolutionary trajectories of the mean alignment strength $\mu_{alg}$ over 700 generations.
        (\textbf{B}) shows the dependence of evolutionary stable strategies (ESS) on the flee strength $\mu_{flee}$. Solid diagonal line shows the theoretically predicted linear dependence of the ESS on $\mu_{flee}$ assuming balancing of social and private information (see SI Sect. C). 
        Dashed lines (\textbf{A}, \textbf{B}) connect the example individual evolutionary runs (\textbf{A}) to the corresponding ESSs (\textbf{B}) obtained as an average over multiple, longer evolutionary simulations.
        (\textbf{C}) Evolutionary stable states (circles) with respect to the group response, measured via the escape ratio $R_{esc}$, for three selected flee-strengths indicated with dashed, solid and dotted lines for $\mu_{flee}=[2, 4, 8]$ respectively.
        In all panels: the dash-dotted magenta line marks the order-disorder transition and the different lines (red, black and blue) represent results for different flee strengths $\mu_{flee}=[2, 4, 8]$, respectively.
    }
    \label{fig:EvoVaryFlee}
\end{figure}

We do find a clear dependence of the ESS on the flee-strength (Fig \ref{fig:EvoVaryFlee}A). More specifically, the ESS exhibits a linear dependence on the flee-strength for $\mu_{flee}\ge 2$ (diagonal line in Fig \ref{fig:EvoVaryFlee}B).
The order transition acts as a lower bound since the non-fleeing agents ($\mu_{flee}=0$) equilibrate closely above it. Thus, the ESS for non-responding agents matches the group-level optimum due to the dynamical school structure at criticality.
 
The linear dependence on the flee-strength may be explained by prey balancing social vs. personal predator information. Social information about the predator is beneficial if the prey is in the second neighbor shell of the predator, i.e. where its neighbors but not itself responses directly to the predator.
Thus, by coordinating with its informed neighbors it gains distance to the predator.
However, if a prey directly senses the predator, social information of uninformed neighbors conflicts with its private information and therefore may hinder evasion. 
Therefore, individual prey agents should continue to evolve towards stronger alignment strength until costs of the social inhibition of evasion counterbalance the benefits of social information. We find support for this conjecture by reproducing the observed linear dependence through a local mean-field approximation (see SI~Sec.~\ref{sec:SI_ESSBalance}) assuming the above balancing mechanism (Fig \ref{fig:EvoVaryFlee}B). Interestingly, also the escape ratio, as a measure of group response while controlling against spatial effects, exhibits a maximum in the strongly ordered region away from criticality (Fig \ref{fig:grpOptima}D).

This leads to the question whether the ESS coincides with the largest escape ratio. Indeed, the maximum of escape ratio shows the same trend as the ESS of moving towards higher alignment strengths with increasing flee strength (Fig \ref{fig:EvoVaryFlee}C), but these maxima stay clearly below the corresponding ESSs (circles in Fig \ref{fig:EvoVaryFlee}C). This suggests that the system does evolve towards unresponsiveness \cite{Torney2015} by increasing the social responsiveness above the optimum (compare column $\mu_{alg}=2$ with $\mu_{alg}=4$ in \ref{mov:4}~Video). We propose that the evolution to unresponsiveness is due to only the targeted prey having a probability of being captured. It appears to be more beneficial for individuals to avoid becoming a target in the first place via a strong social response to fleeing neighbors, rather than being better at escaping once they end up as direct predator targets. Please note, if prey would ignore others during their escape, there would be no trade-off  between social and private information about the predator and agents would remain responsive to the predator at the ESS.

\subsection*{Robustness analysis}

The qualitative results are independent of model implementation details.
We checked for robustness against the predator attack scheme (more and less agile predator), prey-modification (variable speed, persistence length, anisotropy of social interactions / blind angle), modifications in evolutionary algorithm (attack-rate, fitness-estimation) and importantly in a heterogeneous environment (see SI~Sec.~\ref{sec:SI_EvoRobust} and Figs.~\ref{fig:SI_robust},~\ref{fig:SI_EvoFrontKills}).
Note that we explicitly confirmed that considering prey with variable speeds, which enables them to accelerate away from the predator, does not change the qualitative results (SI~Sec.~\ref{sec:SI_robustPrey}). For strong flee forces corresponding accelerations resemble a typical startle response in fish (\ref{mov:5}~Video).

Only by introducing an additional selection pressure, creating a heterogeneous environment, which favors disordered shoals and increasing its weight the ESS may be shifted into the disordered phase. However, even in this case the critical point acts as an unstable evolutionary point (Fig.~\ref{fig:SI_EvoFrontKills}).

Note that our findings are expected to be robust because they are based on generic, model-independent mechanisms: (i) the maximal self-sorting at the transition combined with the spatial explicit implementation of the predator avoidance (causing the transition to be evolutionary unstable) and (ii) the trade-off between social and personal information (causing the ESS to shift to larger social attention with increasing flee strength).
It may be argued that the latter mechanism is biologically not plausible, because prey agents that detect the predator should just flee and ignore their conspecifics. However, this would correspond to a limiting case of a dominating flee-strength and would result in an ESS even further away from the critical point in the highly ordered state (Fig \ref{fig:EvoVaryFlee} B).

\section{Discussion}
We have shown, using a spatially-explicit agent-based model of predator-prey dynamics, that the group optimum with respect to predation avoidance is located in the vicinity of the critical point between disordered swarming and ordered schooling, in line with the so-called ``criticality hypothesis''. However, this optimality is not due to optimal transfer of social information but rather due to the highly dynamical structure of the group at the transition.
Yet, this group optimum at criticality does not represent an evolutionary stable state of individual-level selection. 

Our work demonstrates the crucial importance of taking into account the self-organized spatial dynamics of animal groups when evaluating potential evolutionary benefits of grouping.
It turns out that the mechanism responsible for the optimal collective performance (minimal capture rate) at the critical point, the highly dynamic and flexible structure of the collective, leads also to the steepest selection gradients in evolutionary dynamics, making the critical point evolutionary unstable.  
Evolution with random mutations enforces heterogeneity which in combination with the spatial symmetry breaking at the transition, results in maximal assortative mixing and self-sorting close to the transition. These effects of self-organized collective behavior play a decisive role for the evolutionary dynamics close to criticality and ``drive'' the ESS out of the transition region towards the aligned state. In our system the ESS is in the strongly ordered phase, which suggests the evolution towards external unresponsiveness by overestimating social information.
Finally, we show that the ESS depends linearly on the flee strength, i.e. local perturbation strength, which can be explained by individual balancing of benefits of social information about the predators approach with the costs of social interactions if the information is directly available.

In contrast to Hidalgo et al. \cite{Hidalgo2014}, the critical state in our model is not evolutionary stable, despite the similar setup: evolving agents which respond to conspecifics and to a changing environment (here the appearance of a predator). This can be explained by crucial differences to our work.
Most importantly, in \cite{Hidalgo2014} each agent in isolation can already evolve to its ``individual'' transition by tuning its own gene regulatory network.
This appears to be essential for a critical point corresponding also to the evolutionary stable state in their information-based fitness framework. In our model, the disorder-order transition is a pure collective effect, i.e. individual agents cannot exhibit any transition behavior by themselves.
Furthermore, at the disorder-order transition, small differences in behavioral parameters translate into systematic differences in the self-organized spatial positioning within the group, which in turn directly impacts the predation threat. This self-sorting \cite{Couzin2002, Hemelrijk2005, JamieWood2010} is maximal just above the transition and includes assortative mixing due to emergence of spatial "subgroups" with strong correlations between behavioral phenotype, spatial location and local school structure, which is potentially of interest in the broader context of collective task distribution and computation in spatially-explicit animal groups.

There is another consequence of the tight coupling between local school structure and individual dynamics: The extent of the collective is largest at the transition because the responsiveness to directional fluctuations is maximal, i.e. local fluctuations induce deviations in the movement of different parts of the school causing the school effectively to expand.
In systems with a one-way influence from structure to dynamics (fixed networks) it is known that at the order-transition structural differences cause the largest dynamic variability \cite{Nykter2008}.
We show here that in a system with additional feedback from the dynamics to the structure, also the structure has the highest variability at the transition, which may have important consequences for collective computations, as it may for example enhance collective gradient sensing \cite{Berdahl2013, Hein2015}.
It shows that interactions on fixed \cite{Vanni2011, Brush2016, Chicoli2016} or randomly rewiring \cite{Torney2015} lattices might miss this functionally highly relevant features of collective behavior.

The general structure of the assumed social interactions (short ranged repulsion, alignment and long range attraction) is supported by experiments \cite{Katz2011a, Calovi2018}. However, in different species the detailed dependence of social interactions on relative positions may differ (see e.g. \cite{Calovi2018}). Here, to  be as general as possible,  we used simple functional forms of social interactions. However, the fundamental mechanisms underlying our results such self-sorting and the structure-dynamic feedback will not depend on a more complex, empirically derived, relative position dependence. Neither should alternative interaction mechanisms affect these findings \cite{Bastien2020, Romanczuk2009, Jhawar2020, Lei2020}.

Our finding suggests that evolutionary adaptations at  individual level are not a general mechanism for self-organization towards criticality.  In principle, one could consider the possibility of multi-level selection \cite{Wilson1975, Wilson1997} as a potential mechanism which could make the system evolve towards the group-level optimum at criticality. However, recent theoretical investigations of models of multi-level selection have shown that social dilemma, i.e. differences between ESSs and group level optima, always emerge for non-negligible individual-level selection even in cases where group-level selection strongly dominates \cite{cooney2019replicator,cooney2020analysis}. Thus even in this biologically implausible scenario for fission-fusion prey schools, multi-level selection by its own appears unable to enforce evolutionary stability of the critical point in predator-prey dynamics.

We do not exclude the general possibility that animal collectives may operate in the vicinity of phase transitions in order to optimize collective computations. However, our results clearly demonstrate the necessity for further research on biologically proximate mechanisms of self-organized criticality in animal groups.
A general, fundamental difficulty is that besides predator evasion there are various ecological contexts and other dimensions of (collective) behavior which will affect individual fitness. Here, by focusing on a dominant selection pressure, namely predation, we neglect other mechanisms, as for example resource exploration and exploitation \cite{Hein2015, Brush2016, Wood2007, Monk2018} whose ESS can also depend on the resource abundance \cite{Brush2016, Wood2007, Monk2018}. This emphasizes the importance to study collective behavior in the wild \cite{Handegard2012, Francisco2020, Hansen2020, Graving2019} to provide more empirical input on actual relevant behavioral mechanisms as well as variability of behavior across different contexts. However, we have shown that even by combining two opposing selection mechanisms (see SI~Sec.~\ref{sec:SI_heteroEnvironment}), which on their own favor ordered or disordered state respectively, the critical point does not correspond to an evolutionary attractor, it remains an evolutionary highly unstable point.

 We focused here on the prominent directional symmetry breaking transition between states which are commonly observed in natural systems of collective behavior (disordered swarm, polarized school). Another possible transition involves the milling state \cite{Calovi2015}, however, the function of the milling state in natural systems is unclear. Experiments suggest that boundary effects are a main reason for emergence of milling behavior in the laboratory \cite{Tunstrom2013}, while milling in predator-prey interactions appears only to occur in the final stages of the hunt when the prey school is confined by multiple predators \cite{Wickham1974}.
 
Recently it was suggested that a transition in the speed relaxation coefficient may represent a functionally relevant critical point in flocking behavior \cite{Bialek2013}. Individuals with lower relaxation constants are less bound to their preferred speed and may gain fitness benefits due their ability to adapt faster to higher speeds of fleeing conspecifics.
Consistent with this hypothesis, guppies (\textit{Poecilia reticulata}) exhibit stronger accelerations in high-predation habitats \cite{Herbert-Read2017}.

Fish also exhibit a reflex-driven escape response, so-called startle, which was recently shown to spreads through fish schools as a behavioral contagion process \cite{Sosna2019, Rosenthal2015}. This suggests that at least in the context of collective predator evasion in fish, another type of a critical point may be highly relevant, which is analogous to the critical threshold in epidemic models. It separates states of non-propagating startle response, with only small localized response of single or few individuals, from avalanche-like dynamics, where a single fish may cause a global startle cascade.   
Even if the prey escape behavior is more complex, the self-sorting that happens before or in between predator attacks is unaffected by it and therefore also our results. Additionally, if a school is continuously pursued by predators, as e.g. in pelagic fish \cite{Herbert-Read2015},  the individual prey are likely to swim at their speed limit at which no further acceleration is possible.

Overall, our study does not reject the general possibility that animal groups manifest critical behavior and that it may be adaptive. However, it highlights importance of identification of biologically plausible proximate mechanisms for self-organization towards - and maintenance of - critical dynamics in animal groups, which account for spatial self-organization and the corresponding ecological niche.

% \begin{multicols}{2}
{\footnotesize

% this solution to reset subsection number is found here: https://tex.stackexchange.com/questions/71162/reset-section-numbering-between-unnumbered-chapters
\setcounter{subsection}{0}
\renewcommand*{\theHsection}{chY.\the\value{section}}
\section*{Methods}
All Model parameters are listed in Tab. \ref{tab:SI_paraSummary}.

\subsection*{Prey model}
A prey agent $i$ moves in 2D with constant velocity $v=v_0$ with directional noise of intensity $D$ \cite{Romanczuk2011} and responds to a combined force $\vec{F}_{i}=\vec{F}_{i, alg} + \vec{F}_{i, d} + \vec{F}_{i, flee}$ by adapting its position $\vec{r}_i$ and heading $\varphi_i$ as
\begin{subequations}\label{eq:drdtdvdt}
\begin{align}
    \frac{d \vec{r}_i(t)}{dt} &= \vec{v}_i(t) \\
    \frac{d\varphi_i(t)}{dt} &= \frac{1}{v_0}\left(  F_{i, \perp}(t) + \sqrt{2 D} \xi(t) \right)
\end{align}
\end{subequations}
with $F_{i, \perp}(t) = \vec{F}_i(t) \cdot \vec{e}_{i, \perp}$ as the combined force along the direction $\vec{e}_{i, \perp}=[-\sin\varphi_i, \cos\varphi_i]$ that is perpendicular to the agent's heading direction and $\xi(t)$ as Gaussian white noise.
The alignment force ($\vec{F}_{i, alg}$) between a focal agent $i$ and all its neighbors $j \in \mathbb{N}_i$ is the averaged velocity difference $\vec{v}_{ji} = \vec{v}_j - \vec{v}_i$ times the alignment strength $\mu_{alg}$.
The distance regulating force (see Fig.~\ref{fig:SI_DistanceForce}, panel A) is  
\begin{align}
\vec{F}_{i, d} = \frac{1}{|\mathbb{N}_i|}\sum_{j \in \mathbb{N}_i} \mu_d \cdot \tanh{(m_d (r_{ji} - r_d))} \cdot \hat{r}_{ji} 
\end{align}
with $\hat{r}_{ji} = (\vec{r}_j - \vec{r}_i) / |\vec{r}_j - \vec{r}_i|$ as direction from agent $i$ to $j$, $r_d$ as preferred distance, $\mu_d$ as strength of the force and $m_d$ as the slope of the change from repulsion (for $r_{ji} < r_d$) to attraction (for $r_{ji} > r_d$).
If a predator $p$ is a neighbor, the agent is repelled ($\vec{F}_{i, flee}$) from it with a flee strength $\mu_{flee}$.

\subsection*{Predator-model}
The predator moves with fixed speed $v_p=2v_0$ according to 
\begin{align}
    \frac{d\varphi_p}{dt} = \frac{1}{v_p} \vec{e}_{p, \perp} \cdot \vec{F}_p 
\end{align}
with $\vec{F}_p$ as the pursuit force. It considers its frontal Voronoi-neighbors $\mathbb{N}_p$ as targets and selects equally likely among them ($p_{select, i} = 1/|\mathbb{N}_p|$). It only attacks one prey at a time.
If the predator launches an attack, with an attack rate $\gamma_a$ (also accounting for handling time), its success probability decreases linear with distance and is zero for distances larger than $r_{catch}$:
\begin{align}
    p_{success,i}= \max\left( \frac{r_{catch} - r_{ip}}{r_{catch}}, 0 \right)\ .
\end{align}
In summary, the probability that a predator successfully catches a targeted agent within a small time window $[t,t+\delta t]$ is
\begin{align}
    p_{catch,i}(t,\delta t) = p_{success,i}(t) p_{select,i}(t) \gamma_{a} \delta t.
\end{align}
The pursuit force, with constant magnitude $\mu_p$, points to a weighted center of mass. Each prey position is weighted by its probability of a successful catch $p_{catch, i}(t, \delta t)$. 

\subsection*{Evolutionary algorithm}
The algorithm consists of three components: fitness estimation, fitness-proportionate-selection and mutation.\\
\noindent(i) The fitness is estimated by running $N_f=76$ independent attack-simulations on the same prey population.
For each simulation the $\gamma_a \cdot T_s$ agents with the largest cumulative $p_{catch}$ are declared as dead.
The fitness of agent $i$ is $f_i = -N_{k, i} + max(N_{k, j}, j)$ with $N_{k, i}$ as the number of simulations in which agent $i$ was captured and $max(N_{k, j}, j)$ is the largest number of deaths among all agents.\\
\noindent(ii) The new generation of $N$ offspring is generated via fitness-proportionate-selection.
Thus, a random offspring has the parameters of the parent $i$ with probability $p_{parent, i} = {f_i}/{\sum_j f_j}$.\\
\noindent(iii) An offspring mutates with probability $\gamma_m$ (mutation rate), by adding a Gaussian random variable with zero mean and standard deviation $\sigma_{m}$ to its alignment strength $\mu_{alg}$.

Steps (i) till (iii) are repeated in each generation. To estimate the ESS we compute for each generation the expected offspring population (without mutation to reduce noise) and define the fitness gradient as the offspring mean parameter from which the current mean parameter is subtracted.
Thus, if the offspring have a larger mean parameter, the fitness gradient is positive and vice versa.
The mean fitness gradient of a certain parameter region is the average of generations within it.
For details see SI~Sec.~\ref{sec:SI_EvoAlgo}.

\subsection*{Quantification of collective behavior}
The inter-individual distance is the distance between prey pairs averaged over all pairs $IID=\langle |\vec{r}_{ij}| \rangle$.
The polarization $\Phi$ is the absolute value of the mean heading direction $\Phi = |\vec{\Phi}| = |\sum_i \vec{u}_i/N|$.
The susceptibility $\chi$ is the response of the polarization to an external field $h$ and can be measured via polarization fluctuations
\begin{align}\label{eq:Suscept2}
 \chi = \frac{\partial \Phi}{\partial h} = N( \langle \Phi^2 \rangle - \langle \Phi \rangle^2)
\end{align}
which is a form of the fluctuation dissipation theorem (see SI~Sec.~\ref{sec:SI_suscept}).
It can be shown that Eq.~\ref{eq:Suscept2} is the same as the correlation of velocity fluctuations $\delta \vec{v}_i = \vec{v}_i - \langle\vec{v}\rangle$ over all possible pairs (with $\langle\vec{v}\rangle=\sum_i\vec{v}_i/N$, see SI~Sec.~\ref{sec:SI_suscept}).
However, in inset of Fig \ref{fig:grpOptima}B we computed the correlation of velocity fluctuations only over neighboring pairs $C(\delta \vec{v}_i, \delta \vec{v}_j)=\sum_{i, j\in \mathbb{N}_i} \delta \vec{v}_i \cdot \delta \vec{v}_j$ because it is directly related to local transfer of social information than the correlation over all, including totally unrelated, prey pairs.

We compare the performance of the fleeing prey to the non-fleeing prey (control) using escape ratio
\begin{align}
    R_{esc} = 1-\frac{\gamma_c}{\gamma_{c, NF}}\ .
\end{align}
It is equal to the difference between the capture rates of non-fleeing and fleeing agents $\gamma_{c, NF} - \gamma_c$ scaled by $\gamma_{c, NF}$.
The normalization of the capture difference by the baseline capture rate of non-fleeing prey $\gamma_{c, NF}$ accounts for potential differences in capture rates due to differences in school structure for different parameters, which are unrelated to the fleeing response.

The self-sorting is quantified via the Pearson correlation coefficient between the alignment parameter $\mu_{i, alg}$ of individual agents and their mean relative location in the collective $\langle r_{i, x}\rangle$ where $x\in\{f, s, d\}$ which stands for front, side and local density respectively.
Agents at the front (back) have the largest (smallest) front-location and at the side (center) have the largest (smallest) side-location.
The local density sorting is the correlation of the agents local density and its alignment strength.
For the detailed computation of the relative locations see SI~Sec.~\ref{sec:SI_selfSort}.
Another, more general, quantification of self-sorting is how assortative the spatial arrangement of individuals with heterogeneous alignment is.
We used the implementation of the assortativity coefficient \cite{Newman2003a} in \textit{igraph} on the interaction network (Voronoi) with the values for each agent corresponding to their alignment strength (see SI~Sec.~\ref{sec:SI_measures} for details).

\subsection*{Data availability}
The code to run the predator prey model is available at github (\url{https://github.com/PaPeK/PredatorPrey}).
% \end{multicols}

\bibliographystyle{unsrt}
\bibliography{library}

% \begin{multicols}{2}

\subsection*{Acknowledgments}
We are grateful to Iain Couzin, Simon Levin, Jessica Flack, and David Krakauer for insightful and inspiring discussions leading to this work. We further acknowledge Ishan Levy for suggesting the investigation of non-responding prey dynamics.

Both authors acknowledge funding by the Deutsche Forschungsgemeinschaft (DFG, German Research Foundation) Emmy Noether Programm - RO 4766/2-1.
P. Romanczuk acknowledges funding by the Deutsche Forschungsgemeinschaft (DFG, German Research Foundation) under Germany's Excellence Strategy – EXC 2002/1 “Science of Intelligence” – project number 390523135.

\subsection*{Author contributions}
P.P.K. and P.R. conceptualized the study and wrote the manuscript. P.P.K. performed the research.

\subsection*{Competing interests}
The authors declare no competing interests.

\subsection*{Additional information}
Correspondence and requests for materials should be addressed to P.R. (pawel.romanczuk@hu-berlin.de).
}
% \end{multicols}

\onecolumn
\newpage
% \resetlinenumber[1]
\setcounter{page}{1}
\begin{center}
{\LARGE SI Appendix
\\
Supplementary Information
\\
``Collective predator evasion: Putting the criticality hypothesis to the test''
}
\\
\bigskip
Pascal P. Klamser\textsuperscript{1,2}, Pawel Romanczuk\textsuperscript{1,2*}
\\
\bigskip
\textbf{1} Department of Biology, Institute for Theoretical Biology, Humboldt‐Universität zu Berlin, 10115 Berlin, Germany
\\
\textbf{2} Bernstein Center for Computational Neuroscience, 10115 Berlin, Germany
\bigskip
\bigskip
\end{center}
% \appendix
%%%%%%%%%%%%%%%%%%%%%%%%%%%%%
% Supplementary Information %
%%%%%%%%%%%%%%%%%%%%%%%%%%%%%
% renew equation-labeling:
% \numberwithin{equation}{section}
\renewcommand{\theequation}{S\arabic{equation}}
% renew equation-labeling:
\renewcommand{\thetable}{S\arabic{table}}
% renew equation-labeling:
\renewcommand\thefigure{S\arabic{figure}}    
\setcounter{equation}{0} 
\setcounter{section}{0} 
\setcounter{figure}{0} 
\renewcommand\thesection{\Roman{section}}    

\section{Model-Description}\label{sec:SI_models}
\subsection{Prey-Agents}
The prey agents are modeled as active Brownian particles with constant speed $v=v_0$ and angular noise \citeSI{SIRomanczuk2011}. The stochastic equations of motion read:
\begin{subequations}\label{eq:SI_drdtdvdt}
\begin{align}
    \frac{d \vec{r}_i(t)}{dt} &= \vec{v}_i(t) \\
    \frac{d\varphi_i(t)}{dt} &= \frac{1}{v_0}\left(  F_{i, \perp}(t) + \sqrt{2 D} \xi(t) \right) \ ,
\end{align}
\end{subequations}
with $F_{i, \perp}(t) =  \vec{F}_i(t) \cdot \vec{e}_{i, \perp} $ being the force acting on agent $i$ projected on the direction perpendicular to the direction of motion $\vec{e}_{i, \perp}$, $D$ being the angular diffusion coefficient and $\xi(t)$ being Gaussian white noise with zero mean and vanishing temporal correlations. For simplicity we omit in the following the explicit time dependence of positions, velocities and forces.

Agents react to their environment by (i) coordinating their direction of motion with their neighbors through an alignment interaction, (ii) by trying to maintain a preferred distance to conspecifics (long-ranged attraction and short-ranged repulsion) and (iii) by a fleeing response (repulsion) from the predator.
The alignment force between a focal agent $i$ and all its neighbors $j \in \mathbb{N}_i$ 
\begin{align}\label{eq:Fali}
    \vec{F}_{i, a} = \frac{1}{|\mathbb{N}_i|}\sum_{j \in \mathbb{N}_i} \mu_{alg} \cdot \vec{v}_{ji}.
\end{align}
acts towards minimizing the velocity difference $\vec{v}_{ji} = \vec{v}_j - \vec{v}_i$ with the alignment strength $\mu_{alg}$.

\begin{figure}[h!]
    \centering
    \includegraphics[width=0.7\textwidth]{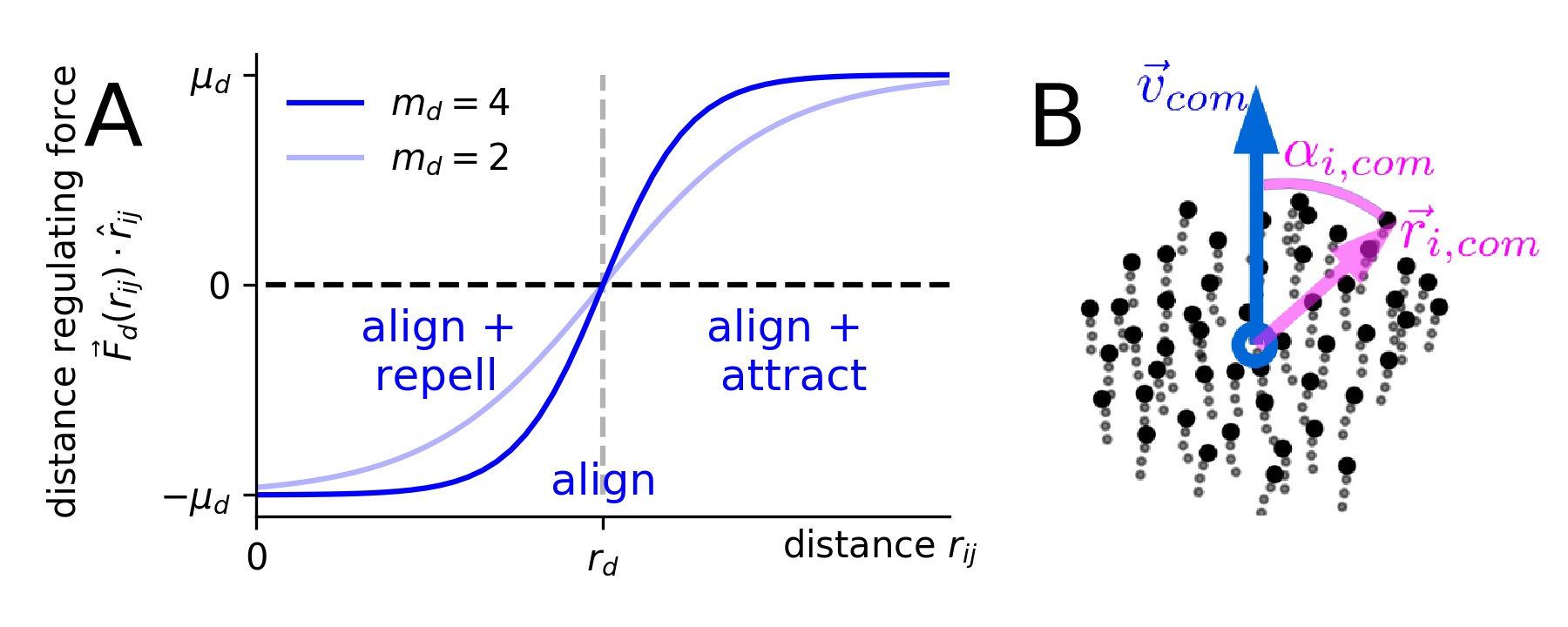}
    \caption{
        {\bf Illustration of the distance regulating force.}
        \textbf{A}: Distance regulating force $\vec{F}_d(r_{ij})$ between agents $i$ and $j$ projected on the separation direction $\hat{r}_{ji}=\frac{\vec{r}_j - \vec{r}_i}{|\vec{r}_j - \vec{r}_i|}$.
        The force equals zero at the preferred distance $r_d=1$ and is displayed for a distance regulating force steepness $m_d=2$ (used in the simulations) and $m_d=4$.
        \textbf{B}: Relative polar coordinates of an agent $i$ with respect to the center of mass $\vec{r}_{com}$ of the school (\textcolor{blue}{blue circle}) and to the average velocity of the school $\vec{v}_{com}$ (\textcolor{blue}{blue arrow}).
        The angle  $\alpha_{i, com}$ (\textcolor{magenta}{magenta arc}) between the school velocity and the agents $i$ current position $\vec{r}_{i, com}$ (\textcolor{magenta}{magenta arrow}) and the distance to the center of mass $|\vec{r}_{i, com}|$ define the position in this relative coordinate system. 
    }
    \label{fig:SI_DistanceForce}
\end{figure}

Individuals attempt to maintain a preferred distance $r_d$ to each other through a distance regulating force  
\begin{align}\label{eq:Fdist}
    \vec{F}_{i, d} = \frac{1}{|\mathbb{N}_i|}\sum_{j \in \mathbb{N}_i} \mu_d \cdot \tanh{(m_d (r_{ji} - r_d))} \cdot \hat{r}_{ji} 
\end{align}
with $\hat{r}_{ji} = \frac{\vec{r}_j - \vec{r}_i}{|\vec{r}_j - \vec{r}_i|}$ being the unit vector along the distance vector from agent $i$ to $j$, $\mu_d$ as strength of the force and $m_d$ as the steepness of the change from repulsion (for $r_{ji} < r_d$) to attraction (for $r_{ji} > r_d$), as illustrated in Fig. \ref{fig:SI_DistanceForce}A.
Finally if a predator $p$ is a neighbor of agent $i$, $p \in \mathbb{N}_i$, the agent is repelled with
\begin{align}
    \vec{F}_{i, f} = - \mu_{flee} \cdot \hat{r}_{pi}
\end{align}
otherwise $\vec{F}_{i,f}=0$.
The total force governing the movement decision of agent $i$ is defined as
\begin{align}
    \vec{F}_i = \vec{F}_{i,d} + \vec{F}_{i, alg} + \vec{F}_{i, flee}\ .
\end{align}

\subsection{Predator-Agent}
For simplicity the predator obeys a deterministic equation of motion for the heading angle, analogous to Eq.~\ref{eq:SI_drdtdvdt}b but without the angular noise term:  
\begin{align}
    \frac{d\varphi_p}{dt} = \frac{1}{v_p} \vec{e}_{p, \perp} \cdot \vec{F}_p\ .
\end{align}
Here, $v_p$ is the fixed predator speed and $\vec{F}_p$ is the predator pursuit force. In this study we consider a predator faster than the prey $v_p > v_0$.
We assume that the predator can only attack one prey at a time. It considers prey individuals which are its frontal Voronoi-neighbors $\mathbb{N}_p$ as targets and selects equally likely among them:
\begin{align} \label{eq:SI_Pselect}
    p_{select, i} = 
    \begin{cases}
        \frac{ 1 }{ |\mathbb{N}_p| }\; &\text{if } i \in \mathbb{N}_p\\
        0 \; &\text{otherwise\ .}
    \end{cases}
\end{align}
The limitation of potential targets to its frontal Voronoi-neighbors $\mathbb{N}_p$, is motivated by kinematic and sensory constraints of the predator.
If the predator launches an attack, with an attack rate $\gamma_a$, which also accounts for potential handling time, it's success probability is linearly dependent on distance and vanishes at distances larger than $r_{catch}$:
\begin{align}
    p_{success,i}=  
        \begin{cases}
            \frac{r_{catch} - r_{ip}}{r_{catch}} & \text{ if } r_{ip} < r_{catch}\\
            0 & \text{ otherwise}.
        \end{cases}
\end{align}
In summary, the probability that a predator successfully catches a targeted agent within a small time window $[t,t+\delta t]$ is
\begin{align}\label{eq:SI_pcatch}
    p_{catch,i}(t,\delta t) = p_{success,i}(t) \cdot p_{select,i}(t) \cdot \gamma_{a} \delta t \ .
\end{align}

The predators movement is biased towards the weighted center of mass of the prey school, where each prey position is weighted by its probability of a successful catch $p_{catch, i}(t, \delta t)$. 
Since $p_{catch, i}$ is non-zero only for the predator's frontal Voronoi-neighbors, the predator movement are governed by local, visually accessible information.
The pursuit force is thus
\begin{align}
   F_p = \mu_p \cdot \left( \sum_i p_{catch, i} \vec{r}_{ip}  \right)\ .
\end{align}

\section{Model parameter}

\begin{table}
    \centering
    \begin{tabular}{c l c c}
         & \textbf{parameter} & \textbf{symbol} & \textbf{value}\\
         \hline
      % \parbox[t]{2mm}{\multirow{3}{*}{\rotatebox[origin=c]{90}{prey}}} 
      \multirow{7}{*}{\rotatebox[origin=c]{90}{\textbf{prey}}} 
         & angular diffusion &      $D$ & 0.5\\
         & alignment strength &     $\mu_{alg}$ & evolves \\
         & distance strength &      $\mu_d$ & 2 \\
         & distance steepness  &        $m_d$ & 2 \\
         & (distance preferred) &       $r_d$ & 1 \\
         & (speed) &                  $v_0$ & 1 \\
         & flee strength &          $\mu_{flee}$ & 4 \\
         \hline
      \multirow{4}{*}{\rotatebox[origin=c]{90}{\textbf{predator}}}
         & speed &                  $v_p$ & 2 \\
         & pursuit strength &       $\mu_p$ &  2 \\
         & attack rate &            $\gamma_a$ & 1/3\\
         & catch radius &         $r_{catch}$ & 3 \\ 
         \hline
      \multirow{4}{*}{\rotatebox[origin=c]{90}{\textbf{simul.}}}
         & number of agents &       $N$ &  400 \\
         & time step &       $dt$ &  0.02 \\
         & equilibration time &       $T_{eq}$ &  200 \\
         & simulation time &       $T_{simu}$ &  120 \\
         & mutation rate &                  $\gamma_m$ & 0.8 \\
         & mutation strength &       $\sigma_m$ &  0.075 \\
    \end{tabular}
    \caption{
        {\bf Default model parameters used.}
        Time and space have been rescaled to dimensionless units by setting, without loss of generality, the prey speed $v_0$ and preferred distance $r_d$ to 1. All length scales are thus measured in units of $r_d$, and all time scales in terms of time needed to move the distance $r_d$.
        Note that the flee strength $\mu_{flee}$ is strictly speaking a predator-prey parameter which reduces the prey-only parameters to four.
    }
    \label{tab:SI_paraSummary}
\end{table}
The default model parameters used are listed in Tab. \ref{tab:SI_paraSummary}.
Note that two parameters can be eliminated by rendering the equations dimensionless.
If, for instance, the preferred distance $r_d$ and the prey speed $v_0$ are used to define the characteristic length $L$ and time $T$:
\begin{align}
    \label{eq:SI_charLandT}
    L = r_d, \; T=\frac{r_d}{v_0},
\end{align}
the Eq.~\ref{eq:SI_drdtdvdt} can be reformulated to
\begin{subequations}\label{eq:SI_drdtdvdtDimLess}
\begin{align}
    \frac{d \vec{r'}_i}{dt'} &= \vec{v'}_i \\
    \frac{d\varphi_i}{dt'} &= \frac{r_d}{v_0^2}\left(  F_{i, \perp} + \sqrt{2 D} \sqrt{\frac{v_0}{r_d}} \xi(t') \right) \\
        &= F_{i, \perp}' + \sqrt{\frac{2 D_{rot} r_d}{v_0}} \xi(t').
\end{align}
\end{subequations}
Here is $D_{rot}=\frac{D}{v_0^2}$ the rotational diffusion coefficient (with the unit $[D]=1/t$). The primed variables are the dimensionless counterparts
\begin{align}
    t = \frac{r_d}{v_0} t',\; v_i=v_0 v'_i,\; r_i=r_d r'_i
\end{align}
and note that the Gaussian stochastic process is transformed according to
\begin{align}
    \xi(t) = \sqrt{\frac{v_0}{r_d}} \xi(t').
\end{align}
With this choice of characteristic length and time and setting $v_0=1$ and $r_d=1$, the dimensionless parameters keep their values listed in Tab. \ref{tab:SI_paraSummary}.

Since the flee strength $\mu_{flee}$ is a predator-prey interaction parameter, the prey system has effectively only four parameters from which the alignment strength $\mu_{alg}$ is evolving. 
The remaining prey parameters are the angular-diffusion coefficient $D$ which is set to $D=0.5$ resulting in a persistence time of $\tau_p=\frac{v_0^2}{D}=2$, i.e. a solitary agents maintains it current direction of motion for approximately the distance of two body length.
The distance regulating strength $\mu_d=2$ is chosen to ensures that the prey group stays cohesive.
The distance steepness $m_d=2$ regulates how quick the distance regulating force saturates to its maximal/minimal values at distances below or above the preferred distance $r_d$ (Fig. \ref{fig:SI_DistanceForce}A).

For the predator the speed must be larger than the prey-speed and is set to $v_p=2$.
Its pursuit strength $\mu_p$ describes together with the speed its turning ability and is set to $\mu_p=2$ and therefore equals the preys distance regulating force strength.
With an capture rate $\gamma_c=1/3$ and a simulation time of $T=120$ around forty prey are captured per round which corresponds to 10\% of the entire school.
The catch radius is set to $r_{catch}=3$ and therefore corresponds to three body length.

The simulation parameters, and in particular the shoal-size of $N=400$, have been chosen in order to simulate biologically reasonable behavior, while at the same time limiting the computational costs. 
For each generation of the evolutionary simulations, 76 independent runs are performed, with each equilibrating for $T_{eq}=200$ before the predator appears, and then running for $T_{simu}=120$ time units.
The time-step is set to $dt=0.02$ which provides sufficient numerical stability and efficient computation (see section\ref{sec:SI_NumericStable}).

\subsection{Numeric stability}\label{sec:SI_NumericStable}
This section addresses the numerical stability of the Euler-Maruyama method used to simulate the stochastic differential equations. The time-step $dt$ should be much smaller than the persistence time $\tau_p=2$, smaller than the shortest correlation time, small enough to fulfill the stability criterion and to avoid oscillating behavior.
An even stricter criterion is that the time step is smaller than a $1/10$ of the correlation time of the fastest process
\begin{align}\label{eq:SI_TenthMaxStochStable}
\frac{1}{10|\mu|} \leq  dt.
\end{align}
Here $\mu$ is the strength of the strongest force  (e.g. alignment-, flee-, repulsion-force).

\section{Evolutionary algorithm and ESS}\label{sec:SI_EvoAlgo}

The evolutionary algorithm is designed to mimic natural selection at the level of behavioral phenotypes.
Among others, the influence of fecundity selection or sexual selection is neglected and the fitness function is only based on how likely an individual is captured in a predator attack, which is a biologically reasonable simplification in the context of predator-prey interactions.
The algorithm consists of (i) a fitness estimation step, (ii) a fitness-proportionate-selection step and (iii) a mutation step.

(i) The fitness is estimated by running $N_f=76$ independent attack-simulations on the same phenotype population.
For each simulation the $\gamma_a \cdot T_{simu}$ agents with the highest cumulative probability of capture (Eq.~\ref{eq:SI_pcatch}) are declared as dead. 
The fitness of agent $i$ is:
\begin{align}
\label{eq:SI_fitEstim}
f_i = -N_{c, i} + max(N_{c, j}, j).
\end{align}
Here $N_{c, i}$ is the number of simulations in which agent $i$ was captured and $max(N_{c, j}, j)$ is the largest number of deaths among all agents.

(ii) The $N$ offspring are generated via the fitness-proportionate-selection.
Thereby has one offspring the parameters of the parent $i$ with probability
\begin{align}
   p_{parent, i} = \frac{f_i}{\sum_j f_j}.
\end{align}

(iii) An offspring agent mutates with a probability $\gamma_m$, the mutation rate, by adding to its alignment strength $\mu_{alg}$ a Gaussian random variable with zero mean and standard deviation $\sigma_{m}$, as the mutation strength.

Steps (i) till (iii) are repeated in each generation.

Note that instead of step (i) the agents could directly get captured during the simulation and removed from the group during the run. This however introduces an additional source of noise in the predation process and the resulting fitness gradient of the prey would become more noisy. As a consequence the number of generations needed to reach an ESS increases.
Nevertheless, to ensure the robustness of our results we repeated the evolution with captures during the evolution, which did not change the final results (see Sect. \ref{sec:SI_EvoRobust}).

\subsection{Estimation of the evolutionary stable state (ESS)}

In the evolutionary algorithm the finite mutation strength and the stochastic roulette-wheel selection introduce noise on top of the intrinsic stochasticity of the the predator-prey dynamics (Eq.~\ref{eq:SI_drdtdvdt}). 
This stochasticity is essential for evolutionary adaptation and exploration of the phenotype space, but makes it challenging to identify the evolutionary stable states (ESS) with high precision in evolutionary simulations.

To circumvent this uncertainty about the exact optimum, we estimate the evolutionary stable state based on the zero-crossing of the fitness-gradient estimated from numerical simulations.
For a system in generation $g$ with agent parameters $\vec{\mu}_{alg}(g)\in \mathbb{R}_+^N$ the estimated fitness gradient $\nabla f(g)$ is computed by predicting the mean outcome of the fitness-proportionate selection
\begin{subequations}\label{eq:SI_predMean}
\begin{align}
    \langle \mu_{alg} \rangle_{predict}(g) &= \vec{p}_{parent, i} \cdot \vec{\mu}_{alg} \\
        &= \frac{1}{\sum_j^N f_j} \sum_i^N f_i \mu_{alg, i}
\end{align}
\end{subequations}
and subtracting from it the current mean-value: 
\begin{align}\label{eq:SI_GenTheoFitGrad}
    \nabla f(g) = \langle \mu_{alg} \rangle_{predict} - \langle \mu_{alg} \rangle.
\end{align}
Note that, in sake of readability, we omitted for terms on the RHS of Eqs. \ref{eq:SI_predMean}, \ref{eq:SI_GenTheoFitGrad} the dependency on the generation $g$.

\begin{figure}
    \includegraphics[width=1\textwidth]{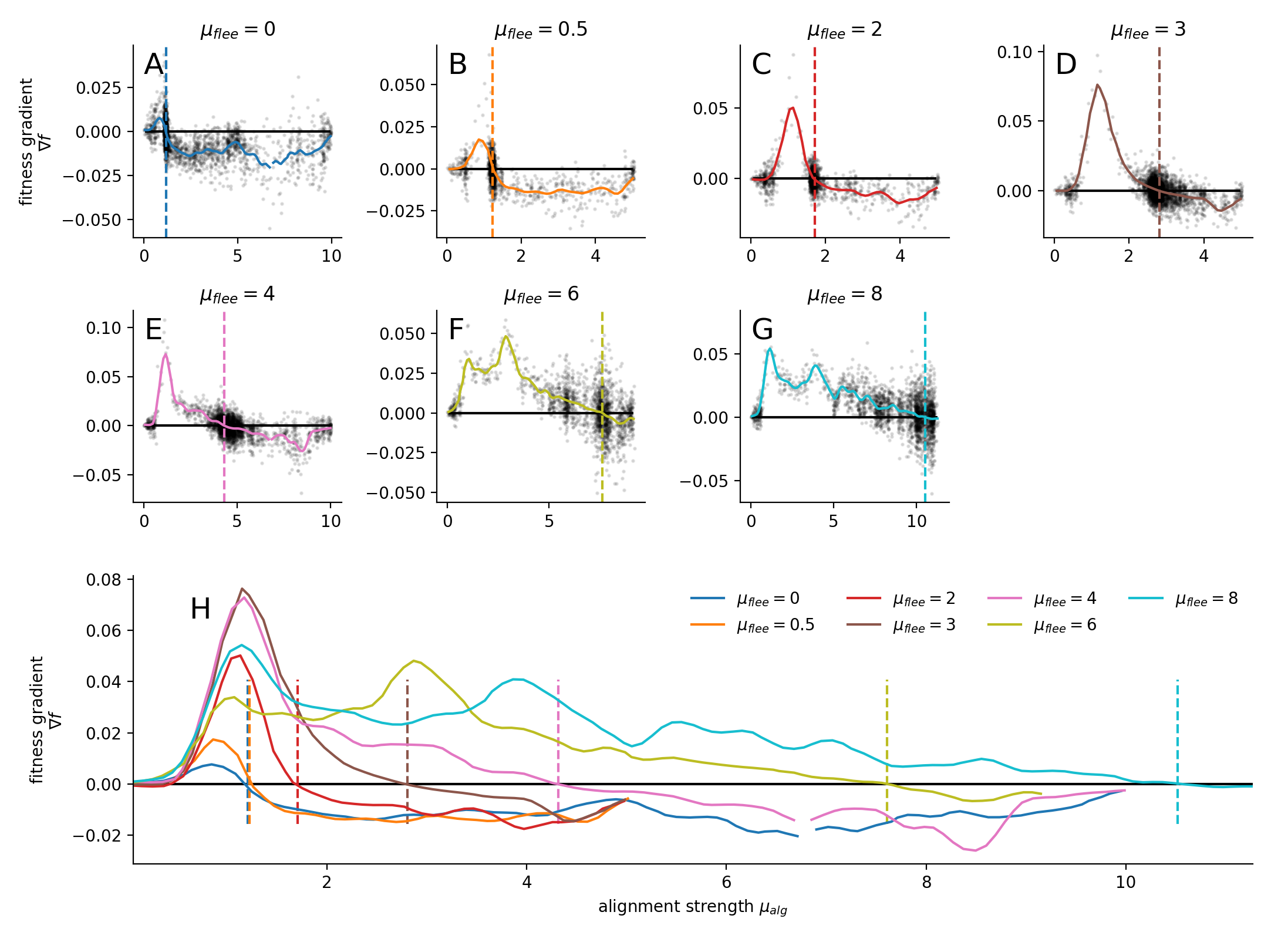}
    \caption{
        {\bf Fitness gradients for different flee-strength to estimate the ESSs.}
        Details on the estimation of evolutionary stable states of Fig. \ref{fig:EvoVaryFlee} in the main text.
        \textbf{A} - \textbf{G}: Fitness gradient $\nabla f$ for evolution with different flee strength $\mu_{flee}$.
        Black-dots indicate the estimated fitness gradients for each generation.
        Solid lines are averaged fitness gradients.
        Dashed vertical lines indicate where $\nabla f = 0$ and thus mark the evolutionary stable states.
        \textbf{H}: All fitness gradients displayed together.
        Note that the peaks for $\mu_{flee}=6$ at $\mu_{alg}\approx 3$ and for $\mu_{flee}=8$ at $\mu_{alg}\approx 4$ are due to fluctuations in the standard-deviation of the population.
        If the standard-deviation is kept constant those peaks vanish (not shown).
    }
    \label{fig:SI_EvoVaryFlee_detail}
\end{figure}

The average fitness gradient corresponding to an alignment strength is
\begin{align}\label{eq:SI_theoFitGrad}
     \nabla f (\mu_{alg}, \Delta_\mu) = 
     \langle \nabla f \rangle_{\mathbb{S}_{\mu_{alg}, \Delta_\mu}} = 
     \frac{\sum_{g\in \mathbb{S}_{\mu_{alg}, \Delta_\mu}} \nabla f(g)}{|\mathbb{S}_{\mu_{alg}, \Delta_\mu}|}
\end{align}
where $\mathbb{S}_{\mu_{alg}, \Delta_\mu}$ is the set of generations which fulfill the condition: 
\begin{align}
\mu_{alg} - \Delta_\mu/2 \leq \langle \mu_{alg} \rangle(g) \leq \mu_{alg} - \Delta_\mu/2.
\end{align}
Therefore, Eq.~\ref{eq:SI_theoFitGrad} represents a simple binning of generations with a bin-width of $\Delta_\mu$.
The maximum of the estimated fitness landscape, i.e. the evolutionary stable state, is where the estimated fitness gradient is zero and where its slope is negative.
An detail illustration of all components needed to compute the ESS as proposed here is shown in Fig. \ref{fig:SI_EvoVaryFlee_detail}.

\section{Measures of self-sorting}\label{sec:SI_measures}
Here we explain in detail the relative positions of individuals in the swarm with respect to the front-back, side-center dimensions and local density.

\subsection{Relative positions}\label{sec:SI_selfSort}

In order to define the relative positions with respect to the front-back and to the side-center dimensions, we first represent every agent position by its distance to the center of mass of the collective
\begin{align}
    r_{i, com} = |\vec{r}_{i, com}| = \vec{r}_{i} - \vec{r}_{com} \ \ \text{with } \vec{r}_{com} = \sum_i \vec{r}_{i} / N
\end{align}
and the angle between its position and the mean velocity of the collective
\begin{align}
    \alpha_{i, com} = \angle(\vec{r}_{i, com}, \vec{v}_{com}) \ \ \text{with } \vec{v}_{com} = \sum_i \vec{v}_i / N\ .
\end{align}
We refer to this representation as the \textit{relative polar coordinates}, illustrated in Fig. \ref{fig:SI_DistanceForce}B.
Note that the x-axis is parallel to $\vec{v}_{com}$, the center of mass is at the origin and the quadrants IV and III are folded onto I and II respectively.
The folding is reasonable if a left-right symmetry holds, which we assume.
The relative front position is
\begin{align}
    \tilde{r}_{i, f} = r_{i, com} \cos{\alpha_{i, com}}
\end{align}
with its normalized version as 
\begin{align}
    r_{i, f} = \frac{ \tilde{r}_{i, f} - \min(\tilde{r}_{j, f}, j) }{\max(\tilde{r}_{j, f}, j) - \min(\tilde{r}_{j, f}, j) }
\end{align}
which results in front positions in the interval $r_{i, f}\in [0, 1]$, with $0$ corresponding to individuals at the very rear of the school and $1$ to individuals at the very front.

The relative side-position is
\begin{align}
    \tilde{r}_{i, s} = r_{i, com} \sin{\alpha_{i, com}}
\end{align}
with its normalized version as 
\begin{align}
    r_{i, s} = \tilde{r}_{i, s} / \max(\tilde{r}_{j, s}, j)\ .
\end{align}

We apply the normalization because we are interested if an individual is at the front and not how far the front is away from the center of mass.
As a results, the normalized measures are less noisy if we average over independent initializations. 
The average normalized relative-position over $S$ samples is 
\begin{align}
    \langle r_{i, x}\rangle = \frac{\sum_{k=1}^{S} r_{i, x, k} }{S}
\end{align}
with $r_{i, x, k}$ as the normalized relative position of agent $i$ in the $k$th sample run.
Note that the normalized relative position is computed after the equilibration time $T_{eq}$.

\subsection{Local density}

The local density of agent $i$ is computed through its distance to the $k$th nearest neighbor $d_{i, kN}$ to
\begin{align}
    \rho_i = k / A(d_{i, kN}, d_{i, e})\ .
\end{align}
The term $A(d_{i, kN}, d_{i, e})$ represents the corrected area.
If the agents distance to the edge of the collective $d_{i, e}$ is larger as $d_{i, kN}$, no correction is needed and the area is the area of a circle with radius $d_{i, kN}$.
If the distance to the edge is smaller than $d_{i, kN}$, the circle-area is corrected by subtracting the area of the circle segment with a sagitta (height) of $h=d_{i, kN} - d_{i, e}$.
Therefore, the area is computed as
\begin{align}
    A(d_{i, kN}, d_{i, e}) =
    \begin{cases}
        \Phi d_{i, kN}^2 &\text{ if } d_{i, kN} < d_{i, e} \\
        \Phi d_{i, kN}^2 - d_{i, kN}\left(d_{i, kN} \arccos \frac{d_{i, e}}{d_{i, kN}} - d_{i, e}\sqrt{1-\frac{d_{i, e}^2}{d_{i, kN}^2}}\right) &\text{ otherwise.}
    \end{cases}
\end{align}
This correction is good if the edge of the collective has a small local curvature compared to the curvature of the circle with radius $d_{i, kN}$.
This should be fulfilled because a collective of $N=400$ individuals with a preferred distance of $r_d=1$ and a spherical form has a radius of $R\approx11$ while the distance to the $k$th
 nearest neighbor with $k=10$ and a Voronoi-interaction network is between 1 and 2.
 
\subsection{Assortativity}
 
 The assortativity $r$ is defined as
\begin{align}
r = \frac{1}{\sigma_q^2} \sum_{j, k} jk (e_{j,k}-q_j q_k) 
\end{align}
with $e_{i,j}$ as the joint probability that a randomly drawn edge connects vertices of type $i$ and $j$, and $q_x$ is the probability that a node of type $x$ is at one end of a randomly drawn edge, i.e. it is the fraction of edges that have a vertex of type $x$ at one end.
The assortativity is the Pearson correlation coefficient over the values of the vertices connected by edges.
 
% CHANGE CHANGE CHANGE CHANGE CHANGE CHANGE CHANGE STARTED
\section{Susceptibility under a homogeneous global field}\label{sec:SI_suscept}
The susceptibility is in general defined by how strong a macroscopic observable $\left< m \right>$ changes if an external field $h$ is changed
\begin{align}\label{eq:suscept}
  \chi = \frac{\partial \left< m \right>}{\partial h}\ .
\end{align}
In the Ising-model, the susceptibility defined in Eq.~\ref{eq:suscept} describes the change of the magnetization per spin 
\begin{align}
m=\frac{M}{N} = \frac{1}{N} \sum_{i=1}^N s_i\ ,
\end{align}
given the change of an external field $h$
The $s_i$ is the spin at side $i$ which can be either up or down, i.e. $s_i \in [-1, 1]$.
Interestingly, the response to a (weak) field can be linked to fluctuations in the order parameter in the absence of a field  \citeSI{SIMarconi2008}.
In statistical physics the probability to observe the system in the state $\vec{s} = [s_0, s_1, \dots, s_N]$ is
\begin{align}\label{eq:probCano}
  P(\vec{s}) = \frac{\exp[-\beta H(\vec{s})]}{Z}\ .
\end{align}
$H(\vec{s})$ describes the energy of the system at state $\vec{s}$ and $\beta$ is the inverse of the thermal energy $\beta=1/(k_b T)$ with $k_b$ as the Boltzmann constant and $T$ as the temperature of the surrounding heat bath.
Thus, the state $\vec{s}$ is more likely the smaller its corresponding energy.
The partition function
\begin{align}\label{eq:partFunc}
  Z=\sum_{\{\vec{s}\}} \exp[-\beta H(\vec{s})]
\end{align}
normalizes the probability with $\sum_{\{\vec{s}\}}$ as a sum over all possible system states.
If spins tend to align with the external field, the energy is partly defined as $H({s_i})=...-h\sum_i s_i$.
Now, the mean magnetization per spin can be computed to
\begin{align}
\left<m\right> = \sum_{\{\vec{s}\}} m(\vec{s}) P(\vec{s}) = 1/N\frac{1}{\beta} \frac{\partial \ln Z}{\partial h}\ .
\end{align}
This allows us to derive the susceptibility $\chi$ defined in Eq.~\ref{eq:suscept} to  
\begin{align}
\label{eq:IsingFDT1}
    \chi = \frac{1}{\beta}\frac{\partial^2 \ln Z}{\partial h^2} = \frac{\beta}{N}[\left< M^2 \right> - \left< M \right>^2] = \beta N[\left< m^2 \right> - \left< m \right>^2] 
\end{align}
The above relation connects the response of the system to an infinitesimally small change of the external field $h$ with fluctuations in the order parameter. The linear nature of this response to small changes can also be assessed by a Taylor-expansion to linear order of the canonical distribution around $h=0$ (see for example Eq.~1.21 in \citeSI{SIMarconi2008}).
The response can be reformulated to highlight the link to the connected spin correlation function or spin pair correlation function
\begin{subequations}
\label{eq:IsingFDT2}
\begin{align}
  \chi &= N \beta [\left< m^2 \right> - \left< m \right>^2] = \frac{\beta}{N} \left[\left< \sum_{ij} s_i s_j \right> - \left< \sum_i s_i \right> \cdot \left< \sum_j s_j \right>\right] \\
      &= \frac{\beta}{N} \sum_{ij}[\left<  s_i s_j \right> - \left< s_i \right> \left< s_j \right>].
\end{align}
\end{subequations}
In the following, we establish an analog description for the model system (presented in Sect.~\ref{sec:SI_models}) with fixed speed.

\subsection{Susceptibility of the prey collective in equilibrium}

For simplicity we assume, as in the section before, that the prey agents (Sect.~\ref{sec:SI_models}) react to a global homogeneous field $\vec{h}$.
From Eq.~\ref{eq:SI_drdtdvdt} the change in heading of individual $i$ in response to $\vec{h}$ is
\begin{align}
   \frac{d \varphi_i}{dt} = \frac{\vec{h}\hat{e}_{\varphi, i}}{v_0} = F_{i,s}\ \text{with}\ \hat{e}_{\varphi, i} = [-\sin\varphi_i, \cos\varphi_i]\ .
\end{align}
From this force $F_{i, s}$ the analog to energy $H_{s, i}$ for individual $i$ can be computed via integration to 
\begin{align}
   H_{s, i} = -\frac{\vec{h}\hat{u}_i}{v_0}\ \text{with}\ \hat{u}_i = [\cos\varphi_i, \sin\varphi_i]\ .
\end{align}
The total energy is composed of the sum of isolated components $H_{s, i}$ and of the part that is influenced by the interactions in between the prey $H_m$:
\begin{align}
H = H_m(\vec{\varphi}) + \sum_i H_{s, i}(\varphi_i, \vec{h}) = H_m(\vec{\varphi}) + -\frac{\vec{h}}{v_0} \cdot \sum_i \hat{u}_i
\end{align}
with $\vec{\varphi}=[\varphi_0, \varphi_1, \dots, \varphi_N]$.
Only $H_{s, i}$ depends on the external field $\vec{h}$.
Knowing the energy of the systems allows (analog to Eq.~\ref{eq:probCano}) to define a probability to observe the state $\vec{\varphi}$ which is
\begin{align}
  P(\vec{\varphi}) = c_H \frac{\exp[\beta \vec{h} \sum \hat{u}_i]}{Z} = c_H \frac{\exp[\beta N \vec{h}\vec{\phi}]}{Z}
\end{align}
with $c_H=e^{-\beta H_m}$.
However, note that Eq.~\ref{eq:probCano} assumes that there is a heat bath represented by $\beta=1/(k_b T)$.
Since the strength of the angular noise $D$ (see Eq.~\ref{eq:SI_drdtdvdt}) can prevent polarization in the prey collective, it plays a similar role as the temperature in the Ising model.
Therefore, we use $\beta=1/(D v_0)$ to compute the expectation value of the polarization vector $\vec{\Phi} = \frac{1}{N} \sum_i^N \hat{u}_i$ (analog to the computation of the mean magnetization in the Ising model).
\begin{subequations}
\begin{align}
    \left< \vec{\Phi} \right>
    &= \sum_{\{\vec{\varphi}\}} \vec{\Phi} P(\vec{\varphi}) = \frac{1}{N \beta} \vec{\nabla}_{\vec{h}} \ln Z \\
    &= \frac{1}{N \beta} \colvec{\frac{\partial}{\partial h_x}}{\frac{\partial}{\partial h_y}}
	\ln\left(\sum_{\{ r, \varphi \}} c_H e^{\beta \vec{h}\cdot \vec{M} }\right)\ ,
\end{align}
\end{subequations}
with $\vec{M}=N\vec{\Phi}$.
Finally, we compute the susceptibility as the sum of changes of the polarization vector $\left< \vec{\Phi} \right>$ components with respect to the external field $\vec{h}$.
It can be written more compact with the $\vec{h}$-Laplace operator $\Delta_{\vec{h}}= \frac{\partial^2}{\partial h_x^2} + \frac{\partial^2}{\partial h_y^2}$ to
\begin{subequations}\label{eq:susceptFromOP}
\begin{align}
   \chi &= \vec{\nabla}_{\vec{h}} \left< \vec{\Phi} \right>
    = \frac{1}{N\beta} \Delta_{\vec{h}} \ln(Z) \\
   &= \frac{\beta}{N} \left[ \left< M_x^2 + M_y^2 \right> - \left( \left< M_x\right>^2 + \left< M_y\right>^2 \right) \right]\\
   &= \frac{\beta}{N} \left[ \left< \vec{M} \cdot \vec{M} \right> - \left< \vec{M} \right> \cdot \left< \vec{M} \right> \right] \\
   &= \beta N \left[ \left< \vec{\Phi} \cdot \vec{\Phi} \right> - \left< \vec{\Phi} \right> \cdot \left< \vec{\Phi} \right> \right]
   = \beta N \left[ \left< \Phi^2 \right> - \left< \Phi \right>^2 \right]\ .
\end{align}
\end{subequations}
This is analogous to Eq.~\ref{eq:IsingFDT2} and establishes a link to the pair-correlation between individual heading direction.
Analogously to Eq.~\ref{eq:IsingFDT2}, we may also write:
\begin{subequations}
\label{eq:susc2corr}
\begin{align}
   \chi &= N \beta \left[ \left<\vec{\Phi} \cdot \vec{\Phi} \right>  - \left< \vec{\Phi} \right> \cdot \left< \vec{\Phi} \right> \right]\\
   &= \frac{\beta}{N} \left[ \left<\sum_i \hat{u}_i \cdot \sum_j \hat{u}_j \right> - N^2 \left< \vec{\Phi} \right> \cdot \left< \vec{\Phi} \right> \right] \\
   &= \frac{\beta}{N} \left[ \left< \sum_{ij} \hat{u}_i \cdot \hat{u}_j \right> - \sum_{ij} \left< \vec{\Phi} \right> \cdot \left< \vec{\Phi} \right> \right] \\
   &= \frac{\beta}{N} \sum_{ij} \left[ \left< \hat{u}_i\cdot \hat{u}_j \right> - \left< \vec{\Phi} \right> \cdot \left< \vec{\Phi} \right> \right] \\
   &= \frac{\beta}{N} \sum_{ij} \left< \left(\hat{u}_i - \left< \vec{\Phi} \right>\right) 
      \cdot \left(\hat{u}_j - \left< \vec{\Phi} \right>\right) \right>\ .
\end{align}
\end{subequations}
Note that the above derivation until Eq.~\ref{eq:susceptFromOP} assumes a thermodynamic equilibrium and is for the out-of-equilibrium prey model strictly speaking not valid (see \citeSI{SISarracino2019, SIMarconi2008} for discussion of non-equilibrium approaches).
However, from Eq.~\ref{eq:susceptFromOP} to Eq.~\ref{eq:susc2corr} there is no such assumption. It is merely a reformulation and therefore valid.
%%%%%%%%%%%%%% PAWEL: THE FOLLOWING SENTENCE IS UNCLEAR %%%%%%%%%%%%%%%%%%%%%%%%%
%%%%%%%%%%%%%% Pascal: I shortened the first and added a rephrased second sentence %%%%%%%%%%%%%%%%%%%%%%%%%
It means, we can interpret $\chi$ always as the sum over the correlation in velocity fluctuations over all pairs. In other words, the larger $\chi$ the stronger is the mean correlation of directional information between random pairs.
%%%%%%%%%%%%%%%%%%%%%%%%%%%%%%%%%%%%%%%%%%%%%%%%%%%%%%%%%%%%%%%%%%%%%%%%%

\subsection{Difference between susceptibility and predator response}\label{sec:cri_suscVsPredResp}

We assumed in Sect.\ref{sec:SI_suscept} that (i) the system is in thermodynamic equilibrium (ii) the changes of the external field are small and it is (iii) global and (iv) homogeneous.
These four are in general violated for the reaction of a collective to a predator.
\begin{itemize}
    \item \textbf{Equilibrium state}: We consider an active system and therefore per definition a non-equilibrium system.
        The agents dissipate constantly energy (no conservation of momentum) but, due to an unspecified energy source, keep their
        preferred speed, i.e. the system is out of thermal equilibrium.
    \item \textbf{Small changes} of an external field: In the context of a predator attack, the perturbing force is 
        the flee-force of the agent.
        This flee-force can also be large and thus can dominate all other forces.
        Therefore, to compute the susceptibility by the linear approximation might not be justified.
    \item \textbf{Global field}: The global homogeneous field simplified the former analytical derivations
        of the susceptibility.  However, the flee-force is neither global nor homogeneous.
        The flee-force acts only on agents that directly sense the predator.
        If we assume visual interactions with occlusion by conspecifics, but also with metric-, Voronoi-interaction and other local interaction types, the predator is per definition a local perturbation.
    \item \textbf{Homogeneous field}:
        The flee-force is in the simplest case a repulsion force and therefore inhomogeneous.
        However, close individuals have similar relative position with respect to the predator and therefore
        also a similar flee-force.
        Thus, locally the force can be approximated to be homogeneous.
\end{itemize}

The violation of the first assumption means that we can not ensure that the fluctuations in the order parameter represent the response of the system to an external field.
However, as shown in Eq.~\ref{eq:susc2corr} these fluctuations are analog with the sum over all pair correlations of velocity fluctuations.
Furthermore, even if we assume that the susceptibility would represent the change of one non-equilibrium stationary state to another one due to an external field, it might be useless at the phase transition.
Phase transitions are up to a certain degree analogous to bifurcations in dynamical systems, i.e. both mark the sudden emergence or extinction of steady states.
Thus, as it is typical for bifurcations, also at phase transitions critical slowing down occurs.
This means that the dynamic of the system slows down and the relaxation to the steady state takes longer the closer the system is to the phase transition.
The attack of a predator is fast and the predator does not wait for the collective to reach a steady state to continue.
This is an additional reason, with the other mentioned unmet assumptions, why the susceptibility should be considered with caution and why its link to optimal predator response is unclear.

% CHANGE CHANGE CHANGE CHANGE CHANGE CHANGE CHANGE ENDED 

\section{Balancing social vs. direct predator information}\label{sec:SI_ESSBalance}
We identified in the main text a possible explanation for the dependence of the evolutionary stable alignment strength on the flee strength as observed in the main text Fig. \ref{fig:EvoVaryFlee}B. % Fig.\ref{fig:EvoVaryFlee} = Fig.3.
A prey can benefit from stronger alignment if it has no private information about the predators position. 
The benefit increases the faster the alignment and therefore should increase with alignment strength.
But if the prey is fleeing already, i.e. it has private (direct) information on the predator position, than alignment to uninformed neighbors can hinder an escape.
Therefore, we expect a balance between benefits and costs.
In the following we will discuss a semi-analytical approximation which reproduces the observed linear dependence.
%Note that we do not want to claim that it is the only possible explanation for the linear dependence but a reasonable one.

\begin{figure}
    \includegraphics[width=0.8\linewidth]{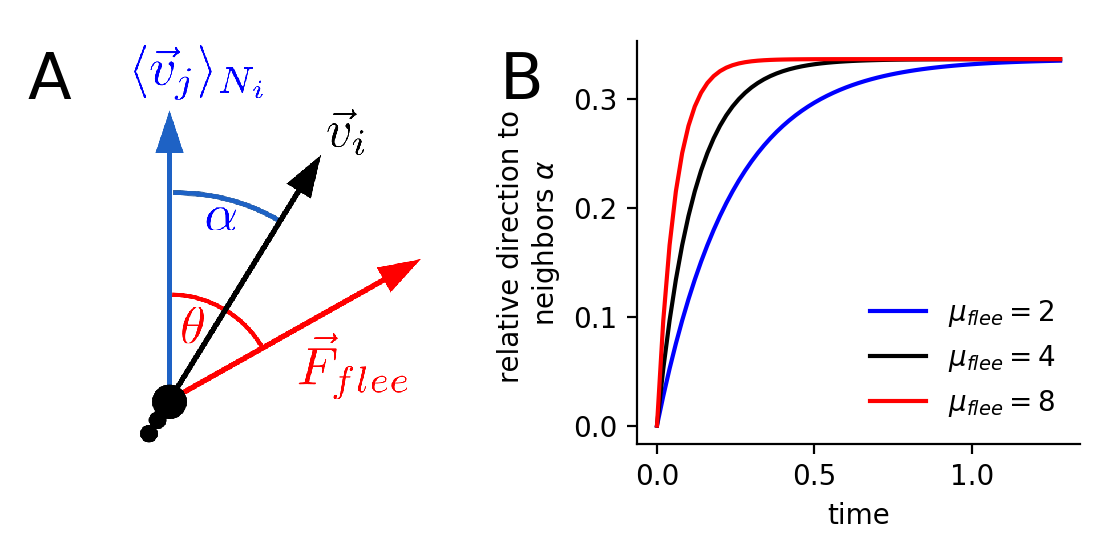}
    \caption{
        {\bf Balancing social and private information via a directional compromise.}
        \textbf{A}: Illustration of angle-vector-relations for variables used in Eq.~\ref{eq:SI_dalphadt} and the following.
        The angle $\alpha$ is the angle between the mean velocity of neighbors $\langle \vec{v}_j \rangle_{\mathbb{N}_i}$ (blue arrow) and the velocity $\vec{v}_i$ of agent $i$ (black arrow).
        The angle $\theta$ is the angle between the mean neighbor-velocity and the flee force $\vec{F}_{flee}$ (red arrow).
        \textbf{B}: Numerical-results of the relative direction to neighbors $\alpha$ using Eq.~\ref{eq:SI_dalphadt}.
        The initial conditions is $\alpha=0$, i.e. the focal agent is perfectly aligned with its neighbors.
        The angle between mean neighbor velocity and flee force is $\theta=\pi/2$.
    }
    \label{fig:SI_CompromiseDirection}
\end{figure}

The costs to align with uninformed prey if the predator position is known can be viewed as a deviation from the flee direction, i.e. the prey relaxes to an effective flee direction which is the compromise between the mean direction of its neighbors and the flee direction Fig. \ref{fig:SI_CompromiseDirection}.

We will use the following assumptions:
\begin{itemize}
    \item i) highly ordered: all neighbors are perfectly aligned with each other.
    \item ii) strong forces: the acting forces are strong such that the agents equilibrate quickly in the direction of the force.
    \item iii) constant forces: the flee-angle and the heading of the neighbors are not changing.
    \item iv) no noise: this will enable us to solve the problem analytically.
\end{itemize}
Consequently the change of the direction-angle of Eq.~\ref{eq:SI_drdtdvdt}b can be reformulated to 
\begin{subequations}
\begin{align}
    \frac{d\varphi_i}{dt} &= \frac{1}{v}\left(  F_{i, \varphi} + \sqrt{2 D} \xi \right) \\
     &\approx \frac{1}{v}\left(  F_{i, \varphi} \right) \\
     &\approx \frac{1}{v}  \left (\mu_{flee} \hat{f}_{flee}
                            + \mu_{alg} [ \langle \vec{v} \rangle_{N_i} - \hat{e}_{r,i}]
                            \right ) \cdot \hat{e}_{\varphi, i}.
\end{align}
\end{subequations}
With $\langle \vec{v}\rangle_{N_i}$ being the mean velocity of all neighbors of agent $i$ and $\hat{e}_{r, i}$ and $\hat{e}_{\varphi, i}$ are its heading and angular direction, respectively.

Without loss of generality we can permanently rotate the system such that $\varphi=0, \forall t$ which simplifies the vector products since $\hat{e}_{r, i} = [1, 0] = \hat{e}_x$ and $\hat{e}_{\varphi, i} = [0, 1] = \hat{e}_y$.
The angle $\alpha$ between $\vec{v}_i$ and $\langle \vec{v}\rangle_{N_i}$ behaves exactly opposite as $\varphi$ (see Fig. \ref{fig:SI_CompromiseDirection}A) and we describe its dynamics instead:
\begin{subequations}\label{eq:SI_dalphadt}
\begin{align}
     \frac{d \alpha}{dt} &= - \frac{d \varphi}{dt} \\
     &\approx  - \frac{1}{v} \left ( \mu_{flee} \hat{f}_{flee}
                             + \mu_{alg} [ \langle \vec{v} \rangle_{N_i} - \hat{e}_x ]
                             \right) \cdot \hat{e}_y\\
     &\approx  - \frac{1}{v} \left( \mu_{flee} f_{flee, y}
                             + \mu_{alg} \langle \vec{v} \rangle_{N_i, y} \right).
\end{align}
\end{subequations}
With $f_{flee, y} = \sin(\theta-\alpha)$ and by assuming perfect order and unit speed the mean velocity of neighbors is $\langle \vec{v} \rangle_{N_i}=1 \begin{pmatrix} \cos(\alpha) \\ \sin(\alpha) \end{pmatrix}$.
Therefore, the change of $\alpha$ simplifies to:
\begin{subequations}
\begin{align}
     \frac{d \alpha}{dt}
     &\approx  - \frac{1}{v} \left( \mu_{flee} \sin(\alpha-\theta)
                             + \mu_{alg} \sin{\alpha} \right) \\
     &\approx  \mu_{flee} \sin(\theta-\alpha)
                             - \mu_{alg} \sin{\alpha}.
\end{align}
\end{subequations}
The fixed points are, as a sanity check, computed for the extreme cases $\mu_{alg}\gg \mu_{flee}$ and $\mu_{flee}\gg \mu_{alg}$ which are $\alpha^\star=0$ and $\alpha^\star=\theta$, respectively.
There exist in general four fixed points from which only one fulfills the criteria $\alpha^\star/\theta \in [0, 1] \forall \left( \mu_{flee}>0,\ \mu_{alg}>0,\ 0<\theta<\pi/2 \right)$ which is:
\begin{align}\label{eq:SI_alphaFix}
    \alpha^\star(\theta_s,\ \mu_{alg},\ \mu_{flee}) &= \arccos
        \frac{\mu_{alg} + \mu_{flee} \cos\theta}{\sqrt{\mu_{alg}^2 + \mu_{flee}^2 + 2 \mu_{alg} \mu_{flee}\cos\theta}}.
\end{align}

Thus $\alpha^\star$ is the effective flee angle with respect to the mean direction of the neighbors.
The closer it is to the flee angle $\theta$ the smaller the cost of being aligned given the knowledge of the predators position.

Now we assume that individuals evolve such that they maintain $\alpha^\star(\theta_s)$ with respect to a specific $\theta_s$.
Thus, if we know the equilibration point $\mu_{alg, evo}^\star(\mu_{flee, evo})$ for the specific flee strength that was used during the evolution $\mu_{flee, evo}$, we can compute the effective flee angle $\alpha^\star(\theta_s,\ \mu_{alg, evo}^\star,\ \mu_{flee, evo}) = \alpha^\star(\theta_s)$.
If we assume that agents evolve such that the balance between alignment benefit and cost, manifested in the effective flee angle, is kept constant, than we can predict the evolutionary stable state $\mu_{alg}^\star$ for a given flee strength by reformulating Eq.~\ref{eq:SI_alphaFix} to
\begin{align}\label{eq:SI_muAlgFix}
    \mu_{alg}^\star &= \frac{\sin(\theta_s-\alpha^\star(\theta_s))}{\sin\alpha^\star(\theta_s)} \mu_{flee}.
\end{align}
The term $\frac{\sin(\theta-\alpha^\star)}{\sin\alpha^\star}$ does not depend on $\theta_s$ which we confirmed numerically.
Thus, the exact choice of $\theta_s$ is irrelevant and $\frac{\sin(\theta-\alpha^\star)}{\sin\alpha^\star}$ is only the slope which connects the origin and the one evolutionary stable state $(\mu_{alg, evo}^\star, \mu_{flee, evo})$ used to compute $\alpha^\star(\theta_s)$ as shown by the blue line in Fig.~\ref{fig:EvoVaryFlee}B.% \ref{fig:EvoVaryFlee} = 3

Note that the equilibrium alignment strength $\mu_{alg}^\star$ above but close to the order transition is systematically lower than its predicted value, as seen for $\mu_{flee}\in\{2, 3, 4\}$ in Fig. \ref{fig:EvoVaryFlee}B. % \ref{fig:EvoVaryFlee} = 3
This can be explained by a small signal due to the low flee strength, because the system relaxes faster the greater the flee strength $\mu_{flee}$ (see Fig. \ref{fig:SI_CompromiseDirection}B).
An alternative explanation is that the spatial selection due to strong self-sorting dominates at the transition.
This explanation is also in agreement with the ESS for low flee strength ($\mu_{flee}=0.5$) being identical to the one with no flee strength at all ($\mu_{flee}=0$).

\section{Robustness against modifications of the prey \& predator dynamics and the selection mechanism}\label{sec:SI_EvoRobust}

\begin{figure}
    \centering
    \includegraphics[width=1\textwidth]{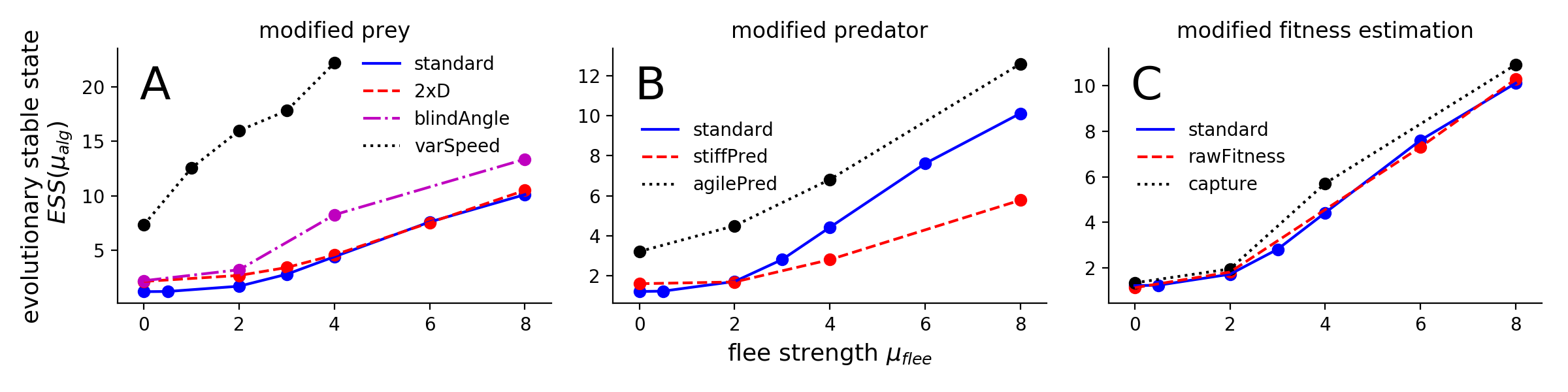}
    \caption{
        \textbf{Robustness analysis of evolution results.}
        Evolutionary stable states of the alignment strength are estimated from the fitness gradient for different flee strength under slight variations of simulations parameters or predator attack implementation.
        The standard scenario of the main text (\textcolor{blue}{blue line}) is compared to
        (\textbf{A}:) a prey population with varying speed which can avoid the predator additionally by acceleration (\textcolor{gray}{black dotted line}), a prey population with a angular diffusion coefficient which is doubled compared to the standard case (\textcolor{red}{red dashed line}), a prey population with a continuous blind angle (\textcolor{magenta}{magenta dash dotted line}),
        (\textbf{B}:) a less agile predator (``stiff'') which turns less quick (\textcolor{gray}{black dotted line}) and an more agile predator which turns quicker (\textcolor{red}{red dashed line}) than the predator in the standard case.
        (\textbf{C}:) a non-binarized fitness estimate (\textcolor  {red}{red dashed line}) in which the prey's fitness is not defined by captures but by the accumulated probability of capture, a fitness estimate based on captures during the simulation (\textcolor{gray}{black dotted line}),
    }
    \label{fig:SI_robust}
\end{figure}

To ensure that our results are robust, we repeat the evolution (Fig. \ref{fig:SI_robust}) with (i) modified prey properties, i.e. changing the angular diffusion coefficient and introducing variable speed and a blind angle, (ii) a changed predator behavior, i.e. its agility, and (iii) changes in the evolutionary selection mechanism, e.g. by an additional high-frontal-risk selection mechanism or by a prey capture during the simulation.
Note that especially the additional high-frontal risk selection is of importance, because it introduces a heterogeneous environment which is assumed to be a general important condition for the evolution to criticality \citeSI{SIHidalgo2014}. 

\subsection{Prey modifications}\label{sec:SI_robustPrey}

The change in angular diffusion from $D=0.5$ to $D=1$ shifts the order-transition to a larger mean alignment strength of $\mu_{alg, c}\approx1.6$ and therefore also increases the lower bound for the ESS which is visible in larger ESS for small flee strength (compare dashed red with blue line in Fig. \ref{fig:SI_robust}A).
For larger flee strength the results are nearly identical suggesting that the mechanism defining the ESS remains unchanged with respect to the standard scenario of the main text.

\begin{figure}
    \centering
    \includegraphics[width=0.8\textwidth]{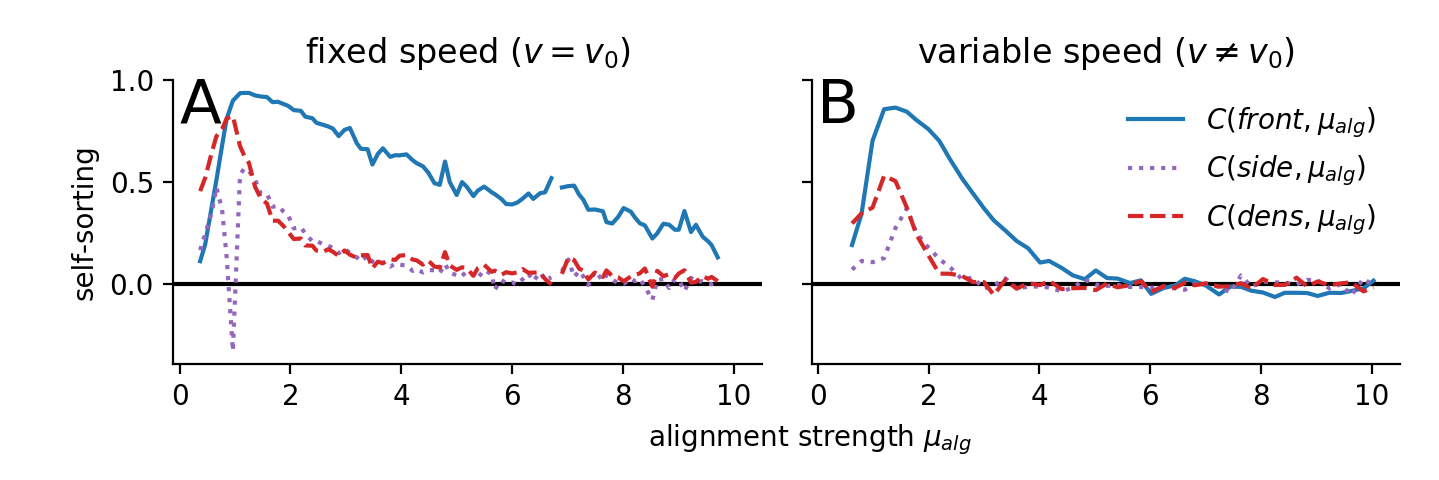}
    \caption{
        \textbf{Self-sorting with and without fixed speed.}
        Self-sorting quantified via the Pearson correlation between the individual alignment parameter $\mu_{alg}$ and the average relative position of the individuals (relative front-, side- or density-location as described in Sect. \ref{sec:SI_selfSort}).
        \textbf{A}: If prey agents respond only by changing their direction but not their speed (fixed speed), self-sorting persists also in highly ordered regions.
        \textbf{B}: If prey agents can change their speed (variable speed), self-sorting vanishes for $\mu_{flee}\leq 6$.
    }
    \label{fig:SI_SelfSortCompare}
\end{figure}

If the speed of the prey is not constant but can change according to social forces, the equations of motion (Eq.~\ref{eq:SI_drdtdvdt}) change to  
\begin{align}\label{eq:SI_drdtVarySpeed}
    \frac{d\vec{r}_i}{dt} &= \vec{v}_i\ \ \text{with}\ \vec{v}_i = v_i [\cos{\varphi_i}, \sin{\varphi_i}]\\
    \frac{dv_i}{dt} &= \beta (v_0 - v_i) + F_{i, v}(t)\\
    \frac{d\varphi_i(t)}{dt} &= \frac{1}{v}\left(  F_{i, \varphi}(t) + \sqrt{2 D} \xi(t) \right)
\end{align}
with $F_{i, v}(t) = \vec{F}_{i} \cdot \hat{e}_{h, i}$ as the projection of the social force of prey $i$ on its heading direction $\hat{e}_{h, i}$ and $\beta$ as the relaxation coefficient which is set in the following to $\beta=4$.
A value of $\beta=4$ prevents the school to relax into a non-moving phase which exists for lower values of $\beta$ \citeSI{SIGrossmann2012}. 
In this non-moving state the speed of the prey would fluctuate around zero.
Additionally, we set an upper bound for the prey's speed corresponding to eighty percent of the predators speed $v_{max}=0.8 v_p$.
Non-fleeing prey ($\mu_{flee}=0$) evolve to significant larger values compared to the standard scenario from the main text (compare dotted black with blue line in Fig. \ref{fig:SI_robust}A).
The ESS for non-fleeing prey ($\mu_{flee}=0$) coincides with the zero-crossing of the front-sorting (Fig. \ref{fig:SI_SelfSortCompare}).
Not only is the ESS of the non-fleeing prey at larger values due to a different self-sorting but also is the ESS much more sensitive to changes in the flee strength (compare slope of dotted black with blue line Fig. \ref{fig:SI_robust}A).
This steeper increase is explainable with an additional social cue, the increased speed of fleeing neighbors, which is not present in the constant speed scenario and goes in hand with findings by Lemmasson et al. \citeSI{SILemasson2009, SILemasson2013}.

% CHANGE CHANGE CHANGE CHANGE CHANGE CHANGE CHANGE STARTED
We introduced an anisotropy of social interactions via a continuous angular preference:
a focal agent $i$ responds stronger to neighbors in front than to those at the side or behind.
Mathematically, the preference depends on the relative angular position $\theta_{ij}$ of neighbor $j\in N_i$, which is the angle between the focal agents current velocity  $\vec{v}_i$ and the relative position of the neighbor $\vec{r}_{ji})$.
Following Calovi et al. \citeSI{SICalovi2014}, the preference decreases with $\theta_{ij}$ 
\begin{align}\label{eq:angPrefer}
  \Omega_{ij} = 1 + \cos\theta_{ij} \text{,  with } \theta_{ij} = \angle(\vec{v}_i, \vec{r}_{ji})\ .
\end{align}
This corresponds to a continuous version of a blind angle.
Thus, instead of computing the social forces by averaging over all Voronoi neighbors equally, a weighted average is performed to compute the alignment and distance regulating force (Eqs. \ref{eq:Fali}, \ref{eq:Fdist}).
The weight is proportional to the angular preference Eq.~\ref{eq:angPrefer}.
This modification leads to less averaging and therefore a higher sensitivity to noise which we measure via a decrease in the polarization for the same parameters as in the standard scenario.
It effectively shifts the disorder-order transition to larger alignment strength (not shown).
In agreement with the shifted disorder-order transition also the ESSs shift to larger alignment strength but the qualitative dependence on the flee strength and their location in the order regime are not altered in comparison to the standard scenario discussed in the main text (compare slope of dash dotted magenta with blue line Fig.~\ref{fig:SI_robust}A).
% CHANGE CHANGE CHANGE CHANGE CHANGE CHANGE CHANGE ENDED 

\subsection{Predator modifications}

We repeated the simulations with (i) a less agile predator which turns slower and (ii) a more agile predator which turns faster compared to the predator considered in the main text.
The different turning ability was implemented by modifying the pursuit strength $\mu_p$ to $\mu_p=1$ for the less agile and to $\mu_p=3$ for the more agile predator.
\\ The effect of using the less agile predator is negligible for low flee-strength, probably because the order-disorder transition acts as lower bound for the ESS due to the explained maximum in assortative mixing and resulting subpopulation selection.
However, for larger flee strength, e.g. $\mu_{flee}\in \{4, 8\}$ in Fig. \ref{fig:SI_robust}B, the ESSs are lowered compared to the standard scenario in the main text.
This can be explained by the missing feedback between the reaction of the prey and the trajectory of the predator:
    in the standard scenario the predator heads for the closest prey, thus if certain prey individuals are good at evading the predator, they have an additional fitness benefit because the predator pursues effectively primarily less well evading prey.
%%%%%%%%%%%%%%%%%%%%%%%%%%% PAWEL: PLEASE CHECK THE FOLLOWING SENTENCE %%%%%%%%%%%%%%%%%%%%%%%%%%%%%%%%%%%%%%%%%%%%%%%%%%%%%%%%%%%%
%%%%%%%%%%%%%%%%%%%%%%%%%%% Pk: Changed. %%%%%%%%%%%%%%%%%%%%%%%%%%%%%%%%%%%%%%%%%%%%%%%%%%%%%%%%%%%%
\\ Consequently, the more agile predator increases the relative fitness benefit of better responding prey and thus amplifies the fitness gradient, which should push the ESS more in the already preferred parameter region.
This is in fact observed (compare dotted black with blue line in Fig. \ref{fig:SI_robust}B).
\\ Despite the quantitative differences due to the predator modifications the general finding discussed in the main text remain unchanged, i.e. that the ESSs are in the ordered phase and increase with increasing flee-strength.
    
\begin{figure}
    \centering
    \includegraphics[width=0.8\textwidth]{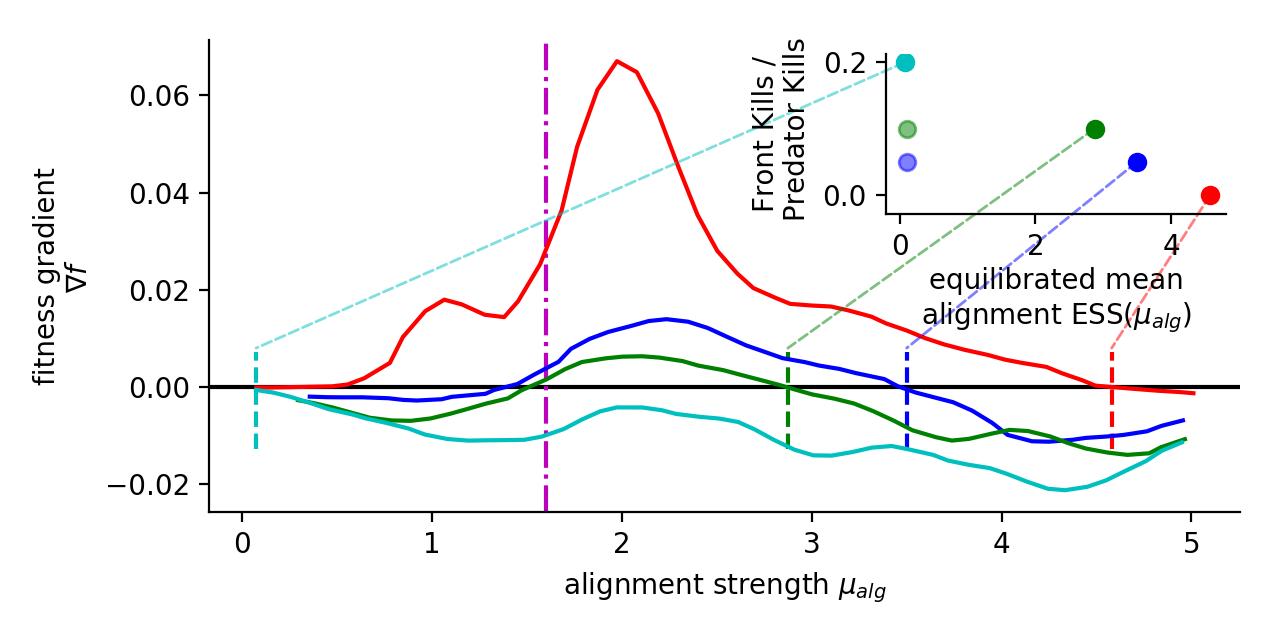}
    \caption{
        \textbf{Evolution in heterogeneous environments.}
        Fitness gradients for different relative strength of the frontal-risk selection with respect to the simultaneously active predator-selection.
        In the frontal-risk selection the most frontal individuals are declared as dead.
        The relative strength of the frontal-risk selection is defined by the ratio between agents killed at the front and by the predator, i.e. $(\text{Front Kills}) / (\text{Pred. Kills}) \in [0,\ 0.05,\ 0.1,\ 0.2]$.
        The evolutionary stable state (ESS) is defined by the zero-crossing of the fitness gradient with negative slope marked by a vertical dashed line.
        However, the lower bound is an additional ESS if the fitness gradient stays negative close to it which is marked by shaded points in the inset.
        Parameters are identical to the former simulations apart from the angular diffusion coefficient which is increased to $D=1$ increasing the order-transition to $\mu_{alg, c}\approx 1.6$ marked by vertical dash-dotted magenta line.
        The flee strength is $\mu_{flee}=4$.
    }
        
    \label{fig:SI_EvoFrontKills}
\end{figure}

\subsection{Selection modification: Evolution in a heterogeneous environment}\label{sec:SI_heteroEnvironment}
In the simulations prey are not captured but a fixed fraction of them with the largest accumulated probability of capture is declared as captured after the simulation.
This means that no prey is removed during the simulation which reduces stochasticity of the fitness estimate but can be considered as unrealistic.
If prey are removed during the simulation based on their current probability of capture and the predators attack rate, the evolution results remain unchanged (compare dotted black with blue line in Fig. \ref{fig:SI_robust}C).
Hereby the attack rate $\gamma_a$ is adjusted at each generation $g$ such that the mean capture rate $\langle \gamma_c \rangle$ matches the initially set attack rate $\gamma_a(g=0)$: 
\begin{align}
    \gamma_a (g+1) = \gamma_a (g) * \frac{\gamma_a(0)}{\langle \gamma_c(g) \rangle}.
\end{align}
This ensures a constant evolutionary pressure.

The attack rate parameter can be abandoned if the fitness is not estimated by the captures but by the negative accumulated probability of capture.
This modification does not alter the ESS identified in the main text at all (compare dashed red with blue line in Fig. \ref{fig:SI_robust}C).

The chosen predator-prey interaction is set as general as possible, nevertheless reasonable alternatives exists and other environmental interactions, e.g. exploration and exploitation of food-sources, might simultaneously impact the fitness. 
We introduce an additional selection mechanisms which favors a disordered phase and creates thus a heterogeneous environment.
The self-sorting for this model predicts that a high mortality of front individuals leads to a disordered state which we implement by declaring the most frontal prey as dead.
This extra selection is equivalent with the observed high risk of being in the front in the presence of sit-and-wait predators \citeSI{SIBumann1997}.
Since the current transition is close to the lower boundary of the alignment parameter ($\min(\mu_{alg})=0$), we set the transition at larger values, i.e. at $\mu_{alg, c} \approx 1.6$,  by increasing the angular diffusion to $D=1$ (ensuring that fluctuations allow equilibration in the disordered regime).

The ESS with respect to alignment decreases with increasing weight on the frontal-risk selection (Fig. \ref{fig:SI_EvoFrontKills}) which seems to be not surprising; however, in a similar study individuals evolved to criticality if exposed to a diverse environment \citeSI{SIHidalgo2014}.
In fact the transition acts here as a fitness valley, marked by a zero-crossing of the fitness gradient with positive slope, causing multiple local optima (inset in Fig. \ref{fig:SI_EvoFrontKills}), which only vanish if one of the selection mechanisms dominates.

\begin{figure}
    \centering
    \includegraphics[width=1\textwidth]{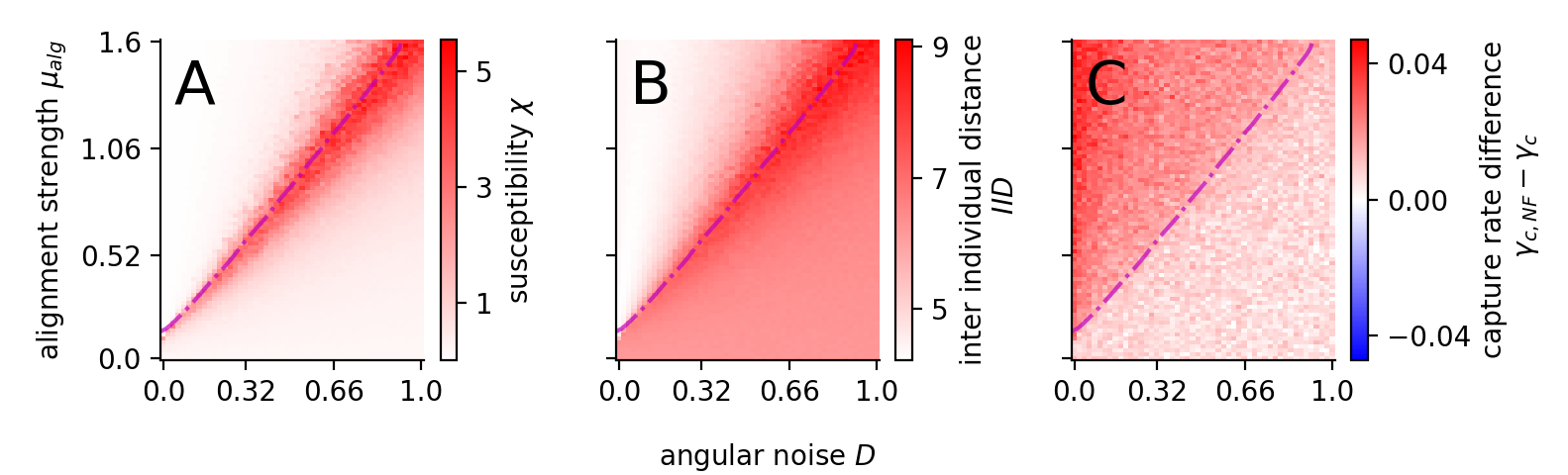}
    \caption{
        \textbf{Enlarged insets from main text Fig. \ref{fig:grpOptima}.}
        The susceptibility $\chi$ (\textbf{A}), inter individual distance IID (\textbf{B}) and difference in capture rate between non-fleeing and fleeing individuals $\gamma_{c, NF} - \gamma_{c}$ (\textbf{C}).
        All measures are shown with colorbars, which were omitted for clarity in Fig. \ref{fig:grpOptima}.
    }
    \label{fig:SI_GrpOptimaInsets}
\end{figure}

\bibliographystyleSI{ieeetr}
\bibliographySI{librarySI}

\newpage
\renewcommand\thesection{\arabic{section}}    
\section*{S Videos}
\setcounter{subsection}{0}
\renewcommand{\thesubsection}{S\arabic{subsection}}

%%%%%%%%%%%%%%%%%%%%%%%% PAWEL: ARE THE TWO FIRST SI MOVIES IDENTICAL? %%%%%%%%%%%%%%%%%%%%%%%%%%%%%%%%%%%%%%%%%%
%%%%%%%%%%%%%%%%%%%%%%%% Pascal: No, 1 Video: without predator, 2 Video: with predator. I made it clearer now (less text :-)) %%%%%%%%%%%%%%%%%%%%%%%%%%%%%%%%%%%%%%%%%%
\subsection{Video}\label{mov:1}
Animation of nine simulations.
The red line are the past- and the empty red circle is the current center of mass of the collective.
Animations in the same column are samples of the same parameter configuration.
The columns differ in the alignment strength $\mu_{alg}=[0, 1, 2]$ indicated at the top.
The remaining parameters are identical to the ones used in the main text (listed in Tab. \ref{tab:SI_paraSummary}).

\subsection{Video}\label{mov:2}
Same as \ref{mov:1}~Video but with a predator attacking the collective.

\subsection{Video}\label{mov:3}
Attack simulation on non- and fleeing prey.
The left panel shows only the fleeing prey, the right the non-fleeing prey, and the center shows both.
The color-code is black=fleeing prey, blue=non-fleeing prey, red=predator attacking fleeing prey, green=predator attacking non-fleeing prey.
Parameters are identical to the ones used in the main text (listed in Tab. \ref{tab:SI_paraSummary}).

\subsection{Video}\label{mov:4}
Same as \ref{mov:2}~Video but with other alignment parameters $\mu_{alg}=[2, 3, 4]$.

\subsection{Video}\label{mov:5}
Animation of nine attack simulations with variable prey speed.
Same as \ref{mov:2} but with preys that are able to accelerate according to the current force. The equations of motions for the prey with variable speed are defined in Sect. \ref{sec:SI_EvoRobust}.

\end{document}